\documentclass[12pt]{article}
\usepackage{graphicx}
\catcode`\@=11
\def\numberbysection{\@addtoreset{equation}{section}
\def\theequation{\thesection.\arabic{equation}}}
\numberbysection

\newcommand{\beq}{\begin{equation}}
\newcommand{\beqa}{\begin{eqnarray}}
\newcommand{\eeq}{\end{equation}}
\newcommand{\eeqa}{\end{eqnarray}}
\renewcommand{\a}{\alpha}
\newcommand{\abs}[1]{\vert#1\vert}
\renewcommand{\d}{{\rm d}}
\newcommand{\de}{_{(2)}}
\newcommand\dis[1]{\displaystyle#1}
\newcommand\ds[1]{\frad{\d#1}{\d s}}
\newcommand\dt[1]{\frad{\d#1}{\d t}}
\newcommand{\e}{{\rm e}}
\newcommand{\evec}{{\bf e}}
\newcommand{\eps}{\varepsilon}
\newcommand{\frad}[2]{\displaystyle{\displaystyle#1\over\displaystyle#2}}
\newcommand{\g}{\gamma}
\newcommand{\gbar}{\overline{g}}
\newcommand{\infy}{_{(\infty)}}
\newcommand{\infymax}{_{(\infty){\rm max}}}
\newcommand{\li}{_{\rm lim}}
\renewcommand{\max}{_{\rm max}}
\newcommand{\m}{{\bf m}}
\newcommand{\mean}[1]{\langle#1\rangle}
\newcommand{\meansur}[1]{\langle\!\langle#1\rangle\!\rangle}
\newcommand{\n}{{\bf n}}
\newcommand{\pl}{$\bullet$}
\newcommand{\s}{\sigma}
\newcommand{\un}{_{(1)}}
\newcommand{\vi}{$\circ$}
\renewcommand{\L}{\Lambda}
\newcommand{\N}{{\cal N}}
\newcommand{\0}{{\bf 0}}
\begin{document}
\centerline{\Large\bf A deterministic model of competitive cluster growth:}
\vspace{.3cm}
\centerline{\Large\bf glassy dynamics, metastability and pattern formation}
\vspace{1cm}

\centerline{\large J.M.~Luck$^{a,}$\footnote{luck@spht.saclay.cea.fr}
and Anita Mehta$^{b,}$\footnote{anita@boson.bose.res.in}}
\vspace{1cm}

\noindent $^a$Service de Physique Th\'eorique\footnote{URA 2306 of CNRS},
CEA Saclay, 91191 Gif-sur-Yvette cedex, France

\noindent $^b$S.N.~Bose National Centre for Basic Sciences, Block JD,
Sector 3, Salt Lake, Calcutta 700098, India
\vspace{1cm}

\begin{abstract}
We investigate a model of interacting clusters which compete for growth.
For a finite assembly of coupled clusters, the largest one always wins,
so that all but this one die out in a finite time.
This scenario of `survival of the biggest'
still holds in the mean-field limit, where the model exhibits
glassy dynamics, with two well separated time scales,
corresponding to individual and collective behaviour.
The survival probability of a cluster
eventually falls off according to the universal law $(\ln t)^{-1/2}$.
Beyond mean field, the dynamics exhibits both aging and metastability,
with a finite fraction of the clusters
surviving forever and forming a non-trivial spatial pattern.
\end{abstract}
\vfill

\noindent P.A.C.S.: 05.45.--a, 47.54.+r, 89.75.--k, 64.60.My.

\newpage
\setcounter{footnote}{0}
\section{Introduction}

Non-equilibrium dynamics can lead to counter-intuitive situations.
One is used, for example, to the premise of equilibration:
in a system with unequally distributed masses,
the effect of most `physical' interactions
would be to bring the system to an equilibrium state
where masses are distributed equally.
This equilibration principle is known to fail in some physical instances,
mostly in the presence of long-range forces,
the prototypical example being gravitational forces.
In fact, as is well known, the effect of gravitation is to amplify forever
the contrasts in mass distribution throughout the Universe~\cite{pee}.

In this work, we interest ourselves in an extreme case of disequilibration.
The model investigated below
deals with immobile interacting clusters which compete for growth.
Although it arose in an astrophysical context,
that of mass accretion by black holes coupled by the radiation field
in a brane world~\cite{archan,I},
its emergent features are relevant to a far wider range of problems.
The present model is strongly out-of-equilibrium
and it obeys no mass conservation law.
A variety of transient behaviour is therefore possible:
for example, two interacting clusters can both decay,
or both grow before one of them dies out.
At late stages, the model follows the {\it survival of the biggest} scenario,
an example of Darwinism in a physical system.
In the mean-field geometry,
the largest cluster generically wins out over all the rest;
in finite dimensions, one has the possibility that infinitely many clusters
survive and grow forever, provided each of them is isolated,
in a sense that will be clearer later on.
This is actually quite meaningful
in the original astrophysical context, since it could, with appropriate
modifications, be adopted to model the scenario of primordial black holes
evolving to a size such that they survive in the present era.

The present model, defined in Section~\ref{model},
deals with an assembly of pointlike, immobile clusters,
which are entirely characterised by their masses.
Cluster masses evolve according to coupled deterministic,
non-linear first-order equations.
We address a medley of situations ranging from finite
assemblies of clusters, to the thermodynamic limit, examined
both in the mean-field geometry and on a lattice
with nearest-neighbour interactions.
Section~\ref{one} describes the dynamical behaviour
of a single isolated cluster:
a large enough cluster, whose initial mass exceeds some threshold,
grows forever, whereas a smaller one evaporates
and disappears in a finite time.
Section~\ref{two} concerns our findings on two interacting clusters,
and more generally finitely many coupled clusters:
the generic scenario is then the survival of the biggest, so that
the largest cluster wins out over all the rest.
The mean-field regime of a large collection of weakly coupled clusters
is investigated in Section~\ref{mft}.
The system exhibits aging and glassy dynamics,
involving two well-separated time scales.
The cluster survival probability decays according to
the universal law $(\ln t)^{-1/2}$.
In Section~\ref{latt}, we examine the model with nearest-neighbour interactions
in finite dimension: the dynamics now exhibits both aging and metastability.
The finite fraction of survivors, i.e., clusters which survive and grow forever,
builds a non-trivial spatial pattern.
In the Discussion (Section~\ref{discussion}),
we put our results in perspective with other growth models.

\section{The model}
\label{model}

The model investigated in this work
is a direct generalisation of that derived in~\cite{archan,I}.
Consider $n$ pointlike, immobile clusters,
which are entirely characterised by their time-dependent masses
$m_i(t)$ for $i=1,\dots,n$.
The cluster masses evolve according to the following
coupled deterministic, first-order equations:
\beq
\dt{m_i}=\left(\frac{\a}{t}-\frac{1}{t^{1/2}}\sum_jg_{ij}\dt{m_j}\right)m_i
-\frac{1}{m_i}.
\label{dtm}
\eeq

These dynamical equations were originally written to model the
kinetics of black hole growth in a radiation fluid~\cite{archan,I}.
In that context, they only hold after some microscopic initial time $t_0$.
The positive (gain) term in the right-hand side of~(\ref{dtm}) represents
mass accretion by the black hole from the surrounding fluid.
The accretion rate in the large parenthesis is the sum of the free rate
for an isolated black hole, proportional to the parameter~$\a>1/2$,
and of the rate induced by all the other black holes via the surrounding fluid.
The coupling $g_{ij}$ between black holes $i$ and $j$
is proportional to the inverse square distance $d_{ij}^2(t_0)$
between them at the initial time $t_0$.
The negative (loss) term in the right-hand side of~(\ref{dtm}) represents
evaporation due to Hawking radiation.

In this paper, the dynamical equations~(\ref{dtm}),
or equivalently~(\ref{dsx}), are now seen as representing
competitive cluster growth.
The functional form of the original equations
derived in~\cite{I} is kept unchanged for definiteness.
The symmetric matrix of couplings $g_{ij}$ reflects the underlying geometry.
Succeeding sections will deal with a raft of scenarios
of ever-increasing complexity, ranging from the dynamics of two coupled
clusters to that of infinitely many.

It turns out to be convenient to switch from physical time $t$ to reduced
(logarithmic) time
\beq
s=\ln\frac{t}{t_0},
\eeq
so that the initial time $t_0$ is mapped onto the origin $s=0$.
Furthermore, we introduce for convenience the reduced masses and square masses:
\beq
x_i=\frac{m_i}{t^{1/2}},\qquad y_i=x_i^2=\frac{m_i^2}{t}.
\eeq
The dynamical equations~(\ref{dtm}) then become the following
{\it autonomous} equations
\beq
\ds{x_i}\equiv x'_i
=\left(\frac{2\a-1}{2}-\sum_jg_{ij}\left(\frac{x_j}{2}+x'_j\right)\right)x_i
-\frac{1}{x_i}
\label{dsx}
\eeq
for the reduced masses $x_i(s)$,
which exhibit no explicit dependence on the reduced time~$s$~\cite{I}.
Throughout the following, accents will denote differentiation with respect
to the reduced time $s$.
It will also be assumed that the couplings~$g_{ij}$
are small enough, so that
\beq
{\rm det}\,(\delta_{ij}+g_{ij}\,x_i)_{i,j=1,\dots,n}>0.
\label{cd}
\eeq
When this inequality holds,
the time derivatives $x'_i$ can be solved explicitly from the
implicit dynamical equations~(\ref{dsx}), so that the dynamics is regular.
Whenever the regularity condition~(\ref{cd}) holds at $s=0$,
it turns out to be preserved by the dynamics.

\section{One isolated cluster}
\label{one}

The simplest situation is that of a single isolated cluster of mass $m(t)$.
The dynamical equation~(\ref{dtm}) reads
\beq
\dt{m}=\frac{\a m}{t}-\frac{1}{m}.
\eeq
The dynamical equation~(\ref{dsx}) simplify to the following ones
for the reduced mass $x(s)$ and square mass $y(s)$:
\beqa
&&x'=\frac{2\a-1}{2}\,x-\frac{1}{x},\\
&&y'=(2\a-1)y-2.\label{yoneds}
\eeqa
Equation~(\ref{yoneds}) is the easier to solve.
It yields at once
\beq
y(s)=y_\star+(y_0-y_\star)\e^{(2\a-1)s},
\label{yone}
\eeq
where $m_0=m(t_0)$ and $y_0=m_0^2/t_0$ are the initial values
of $m(t)$ and $y(t)$, respectively, whereas
\beq
y_\star=\frac{2}{2\a-1}
\label{ystar}
\eeq
is the unstable fixed point of~(\ref{yoneds}).

Returning to physical variables,~(\ref{yone}) reads
\beq
m(t)^2=y_\star t+(m_0^2-y_\star t_0)\left(\frac{t}{t_0}\right)^{2\a}.
\label{m1}
\eeq
This implies that for all $\a>1/2$, we have two kinds of behaviour:

\begin{itemize}

\item
Large clusters, whose initial mass is such that $y_0>y_\star$,
i.e., $m_0$ is larger than the mass threshold
\beq
m_\star=(y_\star t_0)^{1/2}=\left(\frac{2t_0}{2\a-1}\right)^{1/2},
\label{thre}
\eeq
are {\it survivors:} they survive and keep on growing forever.
Equation~(\ref{m1}) reads alternatively
\beq
m(t)^2=m_\star^2\,\frac{t}{t_0}
+(m_0^2-m_\star^2)\left(\frac{t}{t_0}\right)^{2\a}.
\eeq
The second term is the leading one at late times, for all $\a>1/2$.

\item
Small clusters, whose initial mass is below the threshold: $y_0<y_\star$,
i.e., $m_0<m_\star$, evaporate and die out in a finite reduced time,
\beq
s(y_0)=\frac{1}{2\a-1}\ln\frac{y_\star}{y_\star-y_0},
\label{sdis}
\eeq
which diverges logarithmically
as the mass threshold $m_\star$ is approached from below.
The corresponding physical time,
\beq
t(y_0)=t_0\left(\frac{y_\star}{y_\star-y_0}\right)^{1/(2\a-1)},
\eeq
diverges as a power law.

\end{itemize}

Even in this simple case of independent clusters, we get an indication
of a Darwinian scenario (where size, i.e., mass replaces fitness):
the biggest clusters survive, while the smaller ones die out.

Consider now a very large assembly of isolated, i.e., non-interacting, clusters,
characterised by the (continuous) probability distribution function $P(y_0)$
of their initial square masses $y_0$.
One of the quantities of most interest is the survival probability $S(s)$,
defined as the fraction of the clusters which have survived
up to reduced time $s$.
Surviving clusters are exactly those whose initial square mass
obeys $y_0>Y(s)$, where the time-dependent threshold $Y(s)$
is the inverse of $s(y_0)$ introduced in~(\ref{sdis}):
\beq
Y(s)=\left(1-\e^{-(2\a-1)s}\right)y_\star.
\label{bigy}
\eeq
The survival probability at time $s$ reads therefore
\beq
S(s)=\int_{Y(s)}^\infty P(y_0)\,\d y_0.
\label{ss}
\eeq

The limit survival probability\footnote{The subscript $(1)$ recalls
that this result holds for isolated clusters.}
$S\un$ is defined as the fraction of survivors,
i.e., clusters which survive and grow forever.
These are the clusters whose initial mass is above the threshold
$y_\star$ introduced in~(\ref{thre}).
We thus obtain
\beq
S\un=\lim_{s\to\infty}S(s)=\int_{\dis{y_\star}}^\infty P(y_0)\,\d y_0.
\label{s1}
\eeq

For simplicity, we shall often consider in the following
an exponential distribution of initial square masses:
\beq
P(y_0)=\mu\,\e^{-\mu y_0}.
\label{initexpo}
\eeq
The result~(\ref{s1}) then reads
\beq
S\un=\e^{-\mu y^\star}.
\label{sunexpo}
\eeq

\section{Two interacting clusters}
\label{two}

Until now, we have considered only independent clusters:
these survive or not, depending on their initial masses.
We now turn to the more interesting situation of interacting clusters.
The form of the interactions was derived in earlier work~\cite{I}
and is such that clusters could `feed on' each other:
thus, smaller clusters disappear faster
as if they were swallowed by the larger ones.
In this section, we explore the details of the simplest possible case,
that of two interacting clusters with masses $m_1(t)$ and $m_2(t)$,
and interaction strength $g=g_{12}=g_{21}$.

We look successively at the special case of equal masses (Section~4.1)
and at the generic case of unequal masses (Section~4.2).
Only in the first case are the two masses able to survive forever,
growing more slowly than if they had been alone.
In the second case, the bigger cluster swallows the smaller one, generically.

The dynamical equations~(\ref{dsx}) read
\beq
\matrix{
x'_1=\left(\frad{2\a-1}{2}-g\left(\frad{x_2}{2}+x'_2\right)\right)x_1
-\frad{1}{x_1},\cr\cr
x'_2=\left(\frad{2\a-1}{2}-g\left(\frad{x_1}{2}+x'_1\right)\right)x_2
-\frad{1}{x_2}.}
\eeq
Solving these equations for the time derivatives, we obtain
\beq
\matrix{
x'_1=\frad{(2\a-1)x_1^2x_2-2x_2+2g(1-\a x_2^2)x_1^2+g^2x_1^3x_2^2}
{2x_1x_2(1-g^2x_1x_2)},\cr\cr
x'_2=\frad{(2\a-1)x_1x_2^2-2x_1+2g(1-\a x_1^2)x_2^2+g^2x_1^2x_2^3}
{2x_1x_2(1-g^2x_1x_2)}.}
\label{dsx2}
\eeq
In the generic situation of two unequal masses, the above dynamical equations
already illustrate the full complexity of the problem.
The regularity condition~(\ref{cd}), which reads
\beq
1-g^2x_1x_2>0
\label{cd2}
\eeq
in the case of two clusters, is needed for the denominators not to vanish.

\subsection{Equal masses}
\label{twoeq}

The main results of this section can be explained physically
in the following way:
since the interactions cause each mass to `feed on' the other,
overly strong interactions will lead to a strongly depletive effect on both,
as a result of which neither survives.
On the other hand, a weakly interacting pair of equal mass clusters can,
provided their masses are above a threshold, survive in gentle symbiosis;
both depletion and accretion keep occurring at comparable rates,
and the pair survive forever.

Consider now two clusters whose masses are equal at the initial time~$t_0$.
This symmetry is clearly preserved by the dynamics.
Let $x(s)$ be the common value of their reduced mass.
Equations~(\ref{dsx2}) simplify to
\beq
x'=\frac{(2\a-1)x^2-2-gx^3}{2x(1+gx)}.
\label{dsx1}
\eeq

The fixed points of the above dynamical equation, given by
\beq
(2\a-1)x^2-2-gx^3=0,
\label{fp}
\eeq
dictate the qualitative features of the dynamics.
There is a critical value of the coupling,
\beq
g_c=\left(\frac{2(2\a-1)^3}{27}\right)^{1/2},
\label{gc}
\eeq
which separates two kinds of behaviour:

\begin{itemize}

\item For a large enough coupling ($g>g_c$),~(\ref{fp}) has no real
positive root, leading to the absence of a physical fixed point.
The reduced mass $x(s)$ of both clusters decreases monotonically
until they simultaneously die out in a finite reduced time.

\item For a small enough coupling ($g<g_c$),~(\ref{fp}) has three real roots,
two of which are positive and correspond to physical fixed points:
\beq
y_\star^{1/2}<x\un\hbox{ (unstable) }<(3y_\star)^{1/2}<x\de\hbox{ (stable)}.
\eeq
Small clusters, such that $x_0<x\un$,
are attracted by $x=0$, so that both disappear in a finite time.
Large clusters, such that $x_0>x\un$, are attracted by $x\de$:
those pairs of clusters are survivors, and their common mass grows as
\beq
m(t)\approx x\de t^{1/2}.
\eeq
Note that this growth rate is slower than that of an isolated
cluster [see~(\ref{m1})].

\end{itemize}

In other words, small masses, no matter what the coupling strength,
die out in a finite time.
For large reduced masses, the role of the coupling strength matters.
For~$g<g_c$, those larger than $x\un$ survive forever,
and grow more slowly than if they had been isolated;
for $g>g_c$, all die eventually.

These results match those of Section~\ref{one}
in the limit of a vanishingly small coupling.
The unstable fixed point has a finite limit $x\un\to y_\star^{1/2}$,
whereas the stable one diverges as $x\de\approx(2\a-1)/g$.
We recall from Section~\ref{one} that $y_\star$ is the threshold
above which independent clusters survive forever.
The unstable fixed point $x\un$,
which likewise separates dying from surviving clusters,
thus reduces to $y_\star$ in the $g\to0$ limit, as it should.
The main new element with respect to the non-interacting limit is the existence
of the stable fixed point $x\de$.
However, its effect is to attract all masses above $x\un$ to itself,
so that here, too, the effective behaviour is unchanged
with respect to the noninteracting case; masses above $x\un$ survive forever.

\begin{figure}[htb]
\begin{center}
\includegraphics[angle=90,width=.6\linewidth]{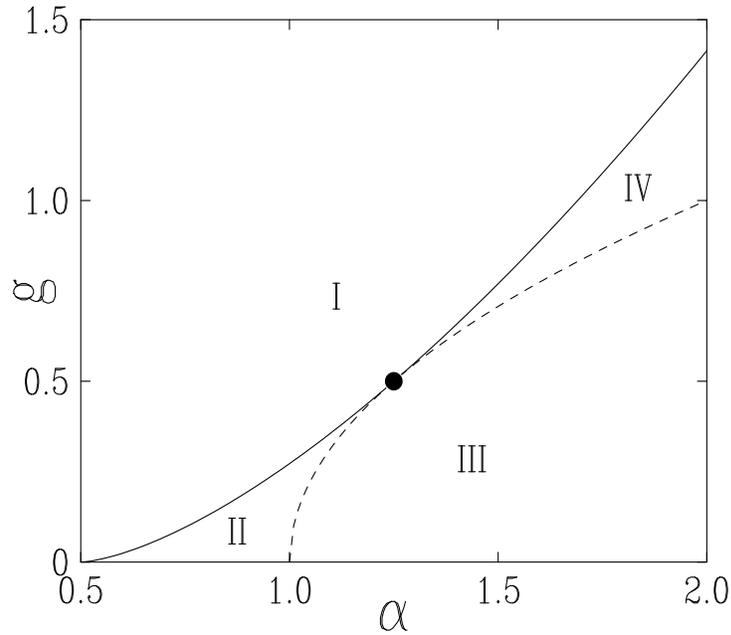}
\caption{\small
Phase diagram of the two-cluster problem in the $\alpha$--$g$ plane.
Full line: critical coupling $g=g_c(\a)$ of~(\ref{gc}).
Dashed line: $g=g\li(\a)$ of~(\ref{gli}).
Phase~I: no real positive fixed point.
Phase~II: two fixed points, both obeying~(\ref{cd1}): $x\un<x\de<x\li$.
Phase~III: two fixed points,
only $x\un$ obeys~(\ref{cd1}): $x\un<x\li<x\de$.
Phase~IV: two fixed points, none obeying~(\ref{cd1}): $x\li<x\un<x\de$.
Full symbol: quadruple point ($\a=5/4$, $g=1/2$): $x\un=x\de=x\li=2$.}
\label{figa}
\end{center}
\end{figure}

Finally, it is worth coming back to the regularity condition~(\ref{cd2}).
In the present situation of two equal masses, this condition reads
\beq
x<x\li,\qquad x\li=\frac{1}{g}.
\label{cd1}
\eeq
The limiting value $x\li$ plays no special role in the dynamics
of two clusters with equal masses.
It will however play an important part
in the transient dynamical behaviour of two unequal masses,
to be studied in Section~4.2.
It is therefore worth investigating the fate of $x\li$.
Equation~(\ref{dsx1}) implies
\beq
x'\vert_{x=x\li}=\frac{\a-1-g^2}{2g}.
\label{lido}
\eeq
This expression singles out the following value of the coupling strength:
\beq
g\li=(\a-1)^{1/2}\qquad(\a>1).
\label{gli}
\eeq
For $g<g\li$, the right-hand side of~(\ref{lido}) is positive,
so that $x\li$ flows toward larger values of $x$.
Conversely, for $g>g\li$, the right-hand side of~(\ref{lido}) is negative,
so that~$x\li$ flows toward smaller values of $x$.
We present in Figure~\ref{figa}, for further reference,
the phase diagram of the two-cluster problem in the $\alpha$--$g$ plane.
Four different phases can be defined (see caption),
according to the number of real positive fixed points,
and to the number of those obeying the condition~(\ref{cd1}).
The phase boundaries are determined by equations~(\ref{gc}) and~(\ref{gli}).

\subsection{Unequal masses}
\label{twoun}

In the general case where the clusters have unequal masses,
the role of the interactions is inherently disequilibrating:
mass differences, however small initially, get rapidly amplified, leading
to the generic scenario of the survival of the biggest.
We shall investigate successively three stages in the dynamics
of two clusters with slightly unequal masses.

\subsubsection*{Early stage: linear stability analysis}

Consider first the early stage of the dynamics
for two clusters with a small mass difference.
Setting
\beq
x_1(s)=x(s)+\eps(s),\qquad x_2(s)=x(s)-\eps(s),
\eeq
to first order in the difference $\eps(s)$,
the mean reduced mass $x(s)$ obeys~(\ref{dsx1}),
while the difference itself obeys the linear equation
\beq
\eps'(s)=\L(x(s))\,\eps(s),
\eeq
where the instantaneous Lyapunov exponent $\L(x)$ reads
\beq
\L(x)=\L_0(x)+\frac{g(2+gx+\a gx^3)}{x(1-g^2x^2)},\qquad
\L_0(x)=\frac{2\a-1}{2}+\frac{1}{x^2}.
\eeq
The full Lyapunov exponent $\L(x)$
of two interacting clusters is therefore larger than
the Lyapunov exponent $\L_0(x)$ in the absence of coupling,
which is in turn larger than the constant $(2\a-1)/2$.
As underlined above, interactions thus enhance disequilibration.
In any case, irrespective of the mean initial mass and of the coupling,
any small initial mass difference diverges
exponentially in the early stages of the dynamics.
In particular, the fixed point $x\de$ of Section~\ref{twoeq},
which is stable against a symmetric perturbation
of the form $\delta x_1=\delta x_2$,
is always linearly unstable against an asymmetric perturbation of the form
$\delta x_1=-\delta x_2=\eps$.

\subsubsection*{Intermediate stage: transient behaviour in the various phases}

Later stages of the dynamics cannot be described in closed form,
because of the non-linearity of~(\ref{dsx2}).
The detailed transient time dependence of both masses depends on the location
of the parameters $\a$ and $g$ in the phase diagram of Figure~\ref{figa},
especially when the initial mass difference is small.

\begin{figure}[htb]
\begin{center}
\includegraphics[angle=90,width=.35\linewidth]{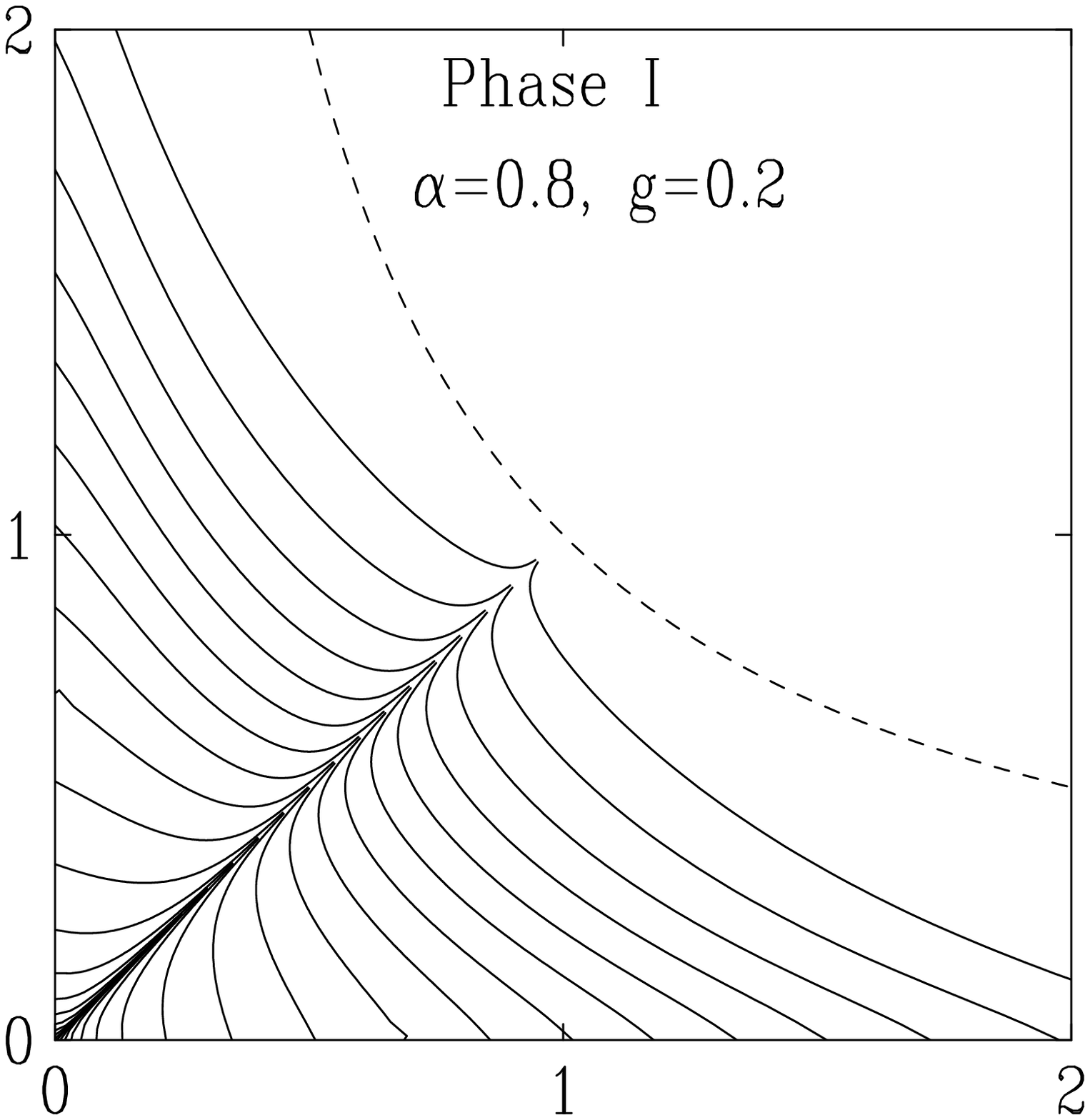}
{\hskip 1mm}
\includegraphics[angle=90,width=.35\linewidth]{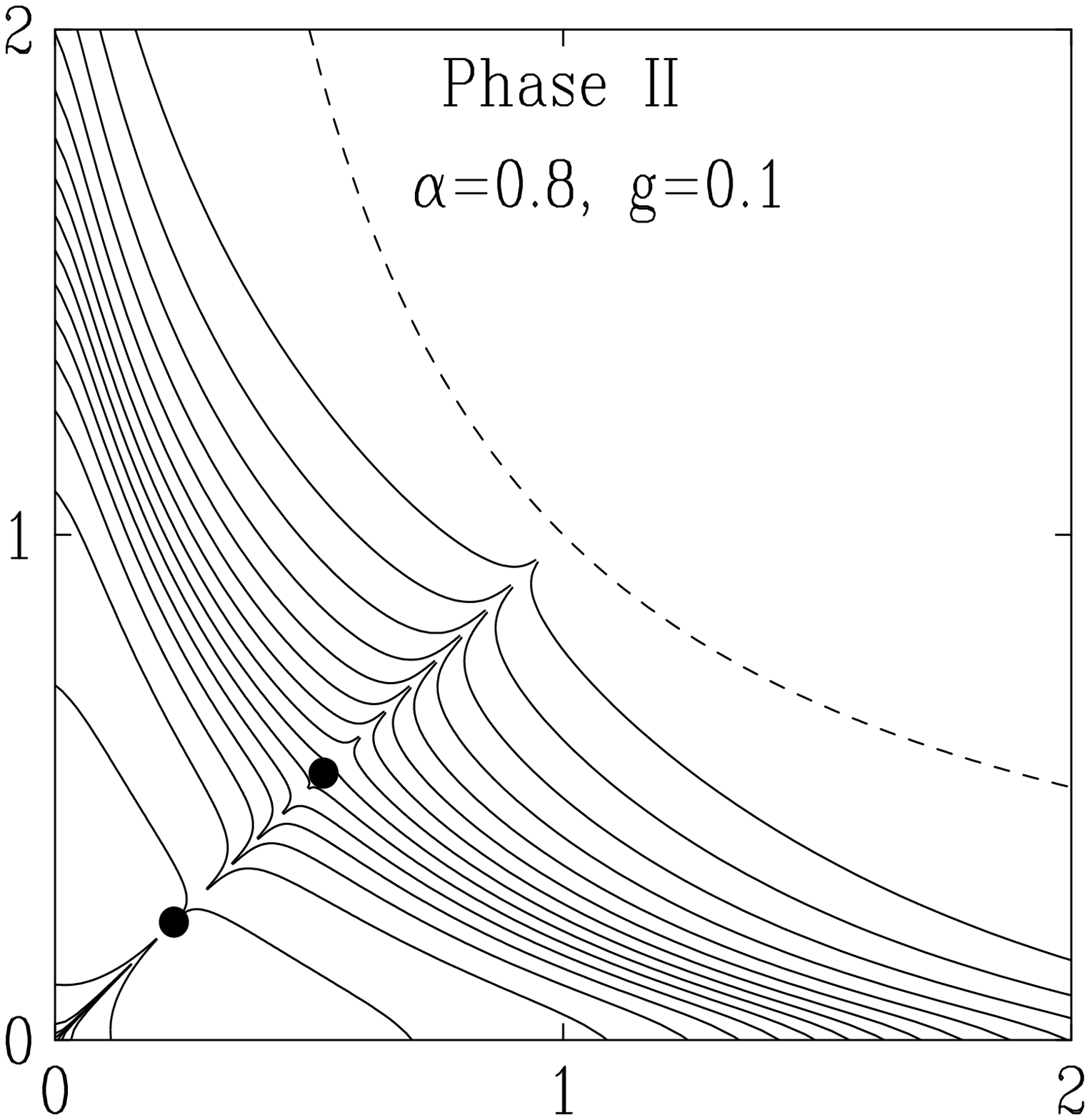}
\vskip 2mm

\includegraphics[angle=90,width=.35\linewidth]{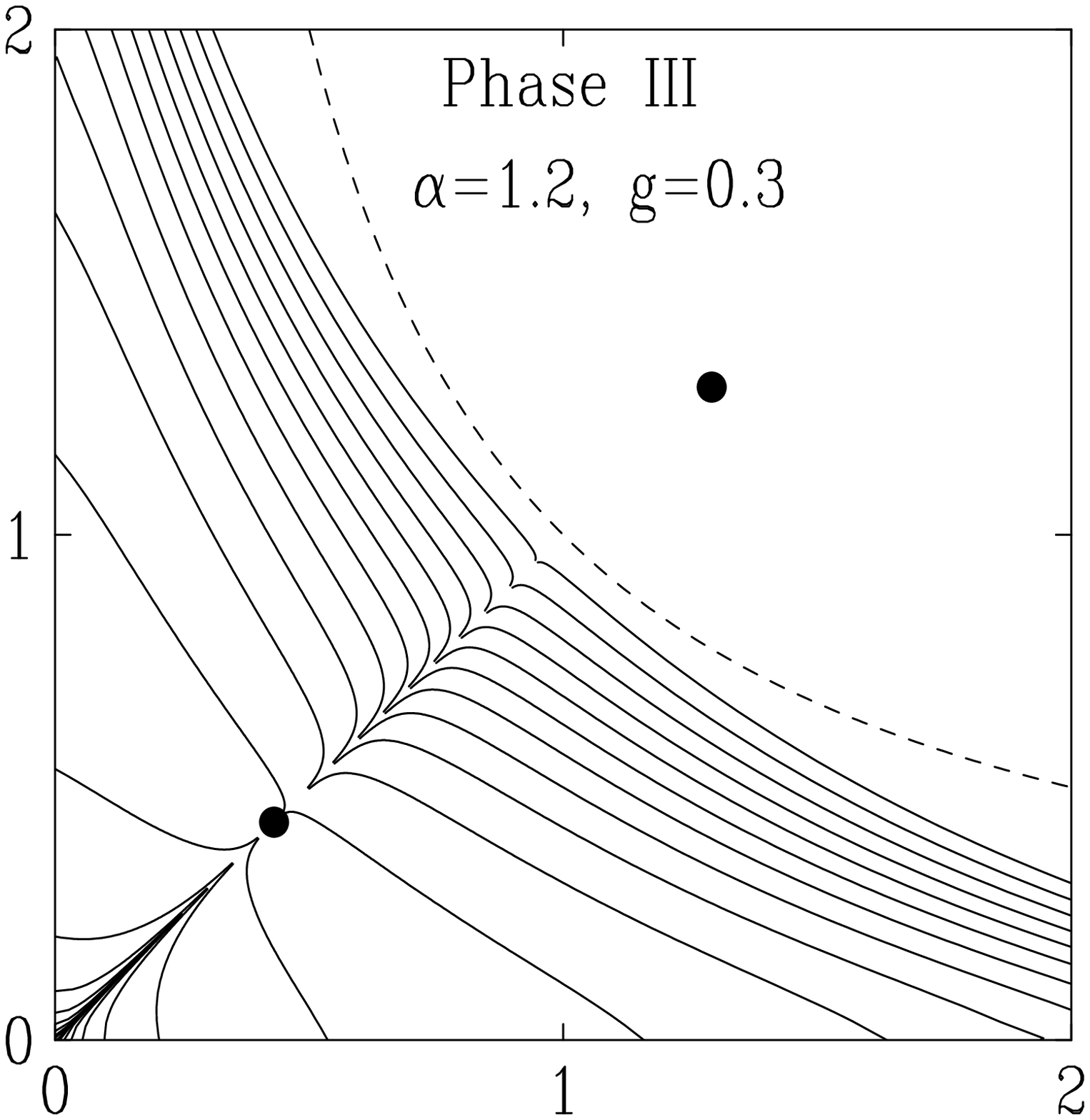}
{\hskip 1mm}
\includegraphics[angle=90,width=.35\linewidth]{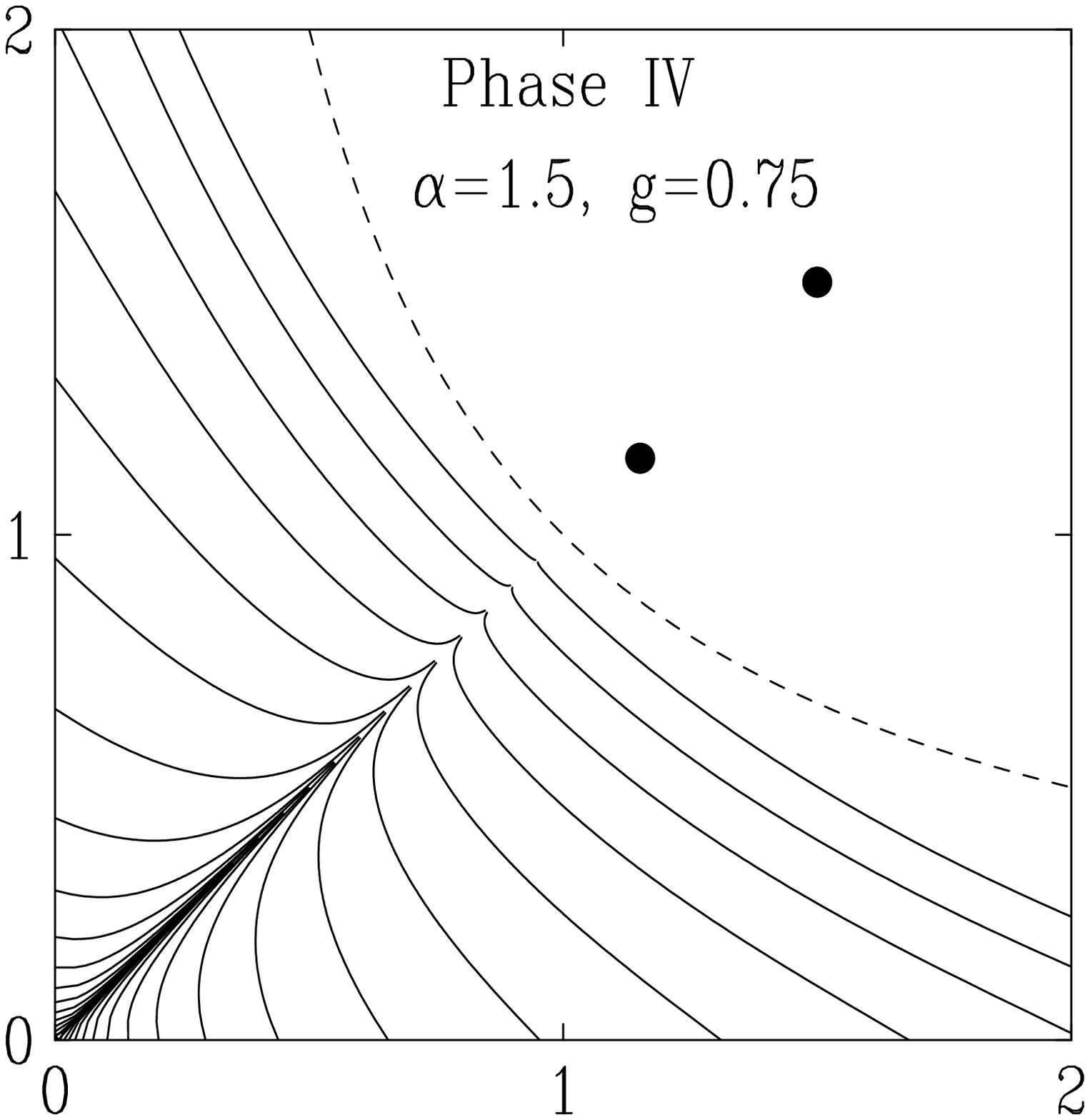}
\caption{\small
Typical trajectories in the $g\,x_1$--$g\,x_2$ plane,
for one choice of the parameters $\a$ and $g$
in each of the four phases shown in Figure~\ref{figa}.
Full lines: a few trajectories with an initial mass difference
$\eps=\pm10^{-2}$.
Dashed line: limit of the regularity condition~(\ref{cd2}),
i.e., $gx_1\cdot gx_2=1$.
Full symbols: fixed points.}
\label{figb}
\end{center}
\end{figure}

These features are illustrated in Figure~\ref{figb},
showing the shape of typical trajectories in the $g\,x_1$--$g\,x_2$ plane.
The dashed line shows the limit of the regularity condition~(\ref{cd2}),
so that allowed pairs of reduced masses are below that line.
Each full line shows a trajectory starting
with a small mass difference $\eps=\pm10^{-2}$.
The mass difference then grows monotonically,
until the trajectory hits either of the co-ordinate axes
in some finite time~$s_1$, when the lighter mass disappears.
Each panel corresponds to a typical choice of $\a$ and $g$
in each of the four phases.
Phases~I and~IV are very similar:
the lighter mass always decreases monotonically,
whereas the larger one decreases in a first stage.
If the larger mass is large enough,
it may then start increasing before the lighter dies out.
Phases~II and~III are also similar:
both masses may increase in a first stage
if their difference is small enough.

\subsubsection*{Late stage: survival of the biggest}

The values of the parameters $\a$ and $g$ become asymptotically
irrelevant in the late stages of the dynamical evolution.
Indeed, as illustrated above,
any trajectory eventually hits either co-ordinate axis.
There is therefore one single generic scenario
for two clusters with unequal initial masses,
i.e., that of survival of the biggest:

\begin{itemize}

\item The smaller one dies out in a finite time $s_1$.

\item The larger one then evolves according to the results of Section~\ref{one}.
Depending on the value of its mass at reduced time $s_1$, it may
either also disappear in a finite time (for $y(s_1)<y_\star$),
or survive and grow forever (for $y(s_1)>y_\star$).

\end{itemize}

The above results still hold qualitatively
for any finite number $n\ge2$ of fully interacting clusters
(all couplings are non-zero).
In the generic situation of unequal masses,
the scenario of survival of the biggest applies:
the $n-1$ smaller clusters die out one after the other,
while only the largest one may become a survivor.

\section{Mean-field limit}
\label{mft}

Having explored the behaviour of finitely many interacting clusters,
we now turn our attention to collective behaviour
in the thermodynamical limit of an infinite assembly of interacting clusters.
In the present section, we focus on the mean-field regime of
long-range interactions.
The main feature of mean-field dynamics is again
that all the clusters eventually die out, except the largest one.
The clusters which survive up to time $s$ are those whose initial reduced
square mass exceeds some time-dependent mass threshold $Y(s)$,
to be determined below.
The general case is necessarily somewhat formal (Section~5.1).
If, however, one considers the regime of weak interactions,
the formalism simplifies considerably (Section~5.2).

\subsection{General formalism}

We consider the mean-field limit of a large assembly of clusters ($n\gg1$),
assuming that all the couplings $g_{ij}$ have the same value $g$.
We perform the usual rescaling of the interaction strength
in mean-field models:
\beq
g=\frac{\gbar}{n}.
\eeq

The problem simplifies drastically in the thermodynamic limit,
defined as usual as the $n\to\infty$ limit at fixed $\gbar$.
In this limit, the coupling strength of any cluster to its whole environment,
measured by $\gbar$, remains of order unity,
whereas the strength of the coupling between any two different clusters,
measured by $g$, falls off as $1/n$.

In this thermodynamic limit,~(\ref{dsx}) implies the following
dynamical equation
\beq
y'(s)=\g(s)y(s)-2
\label{mfds}
\eeq
for the reduced square mass $y(s)$ of any of the clusters.
We have introduced the notation
\beq
\g(s)=2\a-1-\gbar\left(M(s)+2M'(s)\right)
\label{gam}
\eeq
for the effective growth rate of the square mass $y(s)$, where
\beq
M(s)=\mean{x}_s
=\mean{y^{1/2}}_s=\lim_{n\to\infty}\frac{1}{n}\sum_i y_i(s)^{1/2}
\label{stat1st}
\eeq
is the mean reduced mass of the clusters at reduced time $s$.

Despite its apparent simplicity,~(\ref{mfds}) is non-trivial,
because of its self-consistency:
its right-hand-side indeed involves $\g(s)$, and therefore $M(s)$,
and therefore the solution of~(\ref{mfds}) itself.
This self-consistent problem can be solved formally as follows.
First, we have on differentiating~(\ref{stat1st})
\beq
M'(s)=\frac{\g(s)}{2}\,M(s)-N(s),
\eeq
with
\beq
N(s)=\mean{y^{-1/2}}_s=\lim_{n\to\infty}\frac{1}{n}
\sum_{i:y_i(s)>0}y_i(s)^{-1/2},
\eeq
where only non-zero values of $y_i$, corresponding to clusters $i$
which survive at time $s$, are involved in the sum.
The effective rate $\g(s)$ can therefore be solved from~(\ref{gam}):
\beq
\g(s)=\frac{2\a-1+\gbar\,(2N(s)-M(s))}{1+\gbar M(s)}.
\label{gam2}
\eeq
On the other hand, a formal integration of~(\ref{mfds}) yields
\beq
y(s)=G(s)\,(y_0-Y(s)),
\label{y0ys}
\eeq
with
\beq
G(s)=\exp\left(\int_0^s\g(u)\,\d u\right),
\qquad Y(s)=2\int_0^s\frac{\d u}{G(u)},
\eeq
so that
\beq
G(s)=\frac{2}{Y'(s)},\qquad\g(s)=-\frac{Y''(s)}{Y'(s)}.
\label{y2}
\eeq

These steps lead to the following picture of the mean-field dynamics.
For a given initial value $y_0=y(0)$ of the reduced square mass,
the solution~(\ref{y0ys}) holds only as long as $y(s)$ is positive,
or, equivalently, $y_0>Y(s)$.
Hence the smaller clusters, with initial square masses $y_0<Y(s)$,
have already disappeared at reduced time $s$,
while larger ones, such that $y_0>Y(s)$,
have their square masses shifted and dilated from $y_0$ to $y(s)$,
according to~(\ref{y0ys}).

This has strong echoes of the case of many non-interacting clusters.
Recall that there existed a mass threshold, also called $Y(s)$
in~(\ref{bigy}), below which all particles had died at time $s$,
and above which they survived.
The quantity~$Y(s)$ in~(\ref{y0ys}) above
generalises this threshold in the presence of a mean-field coupling.
Now, clusters below this threshold die as before, while the mass
of a cluster grows, as a result of interactions, from $y_0$ to $y(s)$.
In the absence of coupling ($\gbar=0$), the present~$Y(s)$ reduces to that
obtained in~(\ref{bigy}).

The above formalism allows us, provided the continuous probability
distribution of initial square masses $P(y_0)$ is known,
to express all the quantities of interest in terms of
a single (so far unknown) dynamical quantity, the threshold $Y(s)$.
In terms of this, $M(s)$ and $N(s)$ read
\beqa
&&M(s)=G(s)^{1/2}
\int_{Y(s)}^\infty(y_0-Y(s))^{1/2}P(y_0)\,\d y_0,\hfill\label{momm}\\
&&N(s)=G(s)^{-1/2}
\int_{Y(s)}^\infty(y_0-Y(s))^{-1/2}P(y_0)\,\d y_0.\hfill
\eeqa

Equation~(\ref{gam2}) then provides a self-consistent
equation for the unknown quantity~$Y(s)$,
as it only involves $Y(s)$ itself and its first and second derivatives.
Of course, the resulting non-linear integro-differential equation
cannot be solved in closed form in general.

To recapitulate, the program for the mean-field solution
of the dynamics of $n\gg1$ clusters is the following:
Equation~(\ref{y0ys}) gives the growth law of any cluster mass
in terms of the time-dependent mass threshold $Y(s)$.
Only clusters above this threshold survive at reduced time $s$,
as the rest have died.
The calculation of $Y(s)$ can be done self-consistently,
at least in principle, from~(\ref{gam2}).
Quantities of interest, such as the mean mass of surviving clusters at time $s$,
can then be calculated via~(\ref{momm}) and similar expressions.

The definition of the survival probability $S(s)$ is
unaltered by the presence of interactions,
and is still given by equation~(\ref{ss}), i.e.,
\beq
S(s)=\int_{Y(s)}^\infty P(y_0)\,\d y_0.
\label{ss2}
\eeq
The mean reduced mass of the surviving clusters reads
\beq
\meansur{x}_s=\meansur{y^{1/2}}_s=\frac{M(s)}{S(s)},
\label{xmean}
\eeq
where $\meansur{\cdots}_s$ denotes the normalised mean
over the clusters surviving at reduced time~$s$.
In general, average quantities computed over all
clusters, dead and alive, must be renormalised by
$S(s)$ in order to get an appropriately normalised
average over surviving clusters at any time $s$.

\subsection{Weak-coupling regime}

The results of Section 5.1 hold for arbitrary values of $\gbar$,
and their formal nature admits of no further simplification.
However, a much greater transparency is achieved
in the regime where the rescaled coupling $\gbar$ is small.

In this regime, the dynamics consists of two successive stages.
In Stage~I, the clusters behave as if they were isolated,
i.e., their masses evolve according to the results of Section~\ref{one}.
This fast stage of the dynamics therefore corresponds to individual behaviour.
The only surviving clusters after Stage~I are
those whose initial masses exceed the threshold~(\ref{thre}).
The effect of interactions sets in at Stage~II.
This slow stage of the dynamics corresponds to collective behaviour.
All but the largest cluster eventually die out during this stage.

The weakly interacting mean-field regime of our model of interacting
clusters therefore exhibits characteristic features
of glassy systems~\cite{glassyrefs}.
These aging phenomena originate in the presence
of two well-separated time scales of fast and slow dynamics,
with a ratio of respective time scales growing as $1/\gbar^2$.
Another striking feature of our model is the universality
of the main asymptotic results in Stage~II dynamics:
the survival probability falls off generically as $(\ln t)^{-1/2}$,
whereas the mean square mass of survivors grows as~$t\ln t$.
As is well known~\cite{glassyrefs},
such logarithmic behaviour is another telltale sign of glassy dynamics.
We describe all of this below in more detail.

\subsubsection*{Stage~I: Fast individual dynamics}

In this first stage, interactions are essentially irrelevant,
and the dynamics is fast.
The mass of each cluster evolves according to Section~\ref{one},
independently of all the others, as if it were isolated.
The survival probability decays from $S(0)=1$
to the plateau value $S\un$ of~(\ref{s1}),
whereas the time-dependent threshold $Y(s)$ of~(\ref{bigy})
increases from $Y(0)=0$ to~$y_\star$ of~(\ref{ystar}).

\subsubsection*{Stage~II: Slow collective dynamics}

In this second stage, the interactions are responsible
for a slow collective dynamics in the weak-coupling regime.

The evolution throughout Stage~II can be described as follows.
Starting from the assumption (to be checked later on)
that the dynamics is slow, we have $\g(s)\ll1$ and $M'(s)\ll M(s)$.
Equation~(\ref{gam}) therefore simplifies to
\beq
M(s)\approx\frac{2\a-1}{\gbar}.
\label{mslow}
\eeq

Now, inserting~(\ref{y2}) and~(\ref{mslow}) in~(\ref{momm}),
we obtain the following closed differential equation
for the unknown $Y(s)$:
\beq
Y'(s)\approx\frac{2\,\gbar^2}{(2\a-1)^2}\,R(Y)^2,
\label{yslow}
\eeq
where the function
\beq
R(Y)=\int_Y^\infty(y_0-Y)^{1/2}P(y_0)\,\d y_0
\eeq
is entirely determined by the initial mass distribution.

The behaviour of the threshold $Y(s)$ throughout Stage~II
is obtained by integrating~(\ref{yslow}),
with an initial value equal to the plateau value $y_\star$ of~(\ref{ystar}):
\beq
\int_{y_\star}^Y\frac{\d y}{R(y)^2}\approx\frac{2\,\gbar^2\,s}{(2\a-1)^2}.
\label{ws}
\eeq
Equation~(\ref{ws}) contains the key to the dynamical behaviour in Stage~II.

First, the characteristic time of the collective dynamics,
\beq
s_c\sim\frac{(2\a-1)^2}{\gbar^2},
\label{scmf}
\eeq
becomes arbitrarily large in the weak-coupling regime ($\gbar\to0$).
The assumption of slow dynamics is thus fully justified.

Then, the analysis of the long-time dynamics goes as follows.
Note that the left-hand side of~(\ref{ws}) diverges as $s\to\infty$.
The only way this can occur is if $R(Y)$ falls off to zero for long times,
i.e., if the threshold $Y(s)$ goes to the maximum possible
initial mass $y\max$, i.e., more formally,
the upper bound of the continuous distribution $P(y_0)$.
The survival probability~$S(s)$ then also falls off to zero for long times.
We can conclude that the whole population of clusters
which survived Stage~I eventually disappears during
Stage~II of the mean-field dynamics.
For a finite mean-field system consisting of $n$ coupled clusters,
at most one of them will survive forever,
as already mentioned at the end of Section 4.2.

Having established the general pattern,
we now specialise to specific distributions of initial masses,
in order to obtain quantitative predictions.
We first consider an exponential distribution of initial masses.
We will find that the results
obtainable from it can be generalised to a raft of other distributions.

With the exponential distribution~(\ref{initexpo}),
equations~(\ref{ss2}),~(\ref{ws}) yield
\beq
\e^{2\mu Y(s)}=\frac{1}{S(s)^2}
\approx\e^{2\mu y_\star}+\frac{\pi\,\gbar^2\,s}{(2\a-1)^2}.
\eeq
In the late times of Stage~II, the survival probability therefore decays as
\beq
S(s)\approx\frac{2\a-1}{\gbar}\,(Cs)^{-1/2},
\label{slate}
\eeq
with
\beq
C=\pi
\label{cpi}
\eeq
for the chosen exponential distribution,
irrespectively of $\a$, $\mu$, and~$\gbar$, provided the latter is small enough.

Equations~(\ref{xmean}),~(\ref{mslow}) then lead to
\beq
\meansur{x}_s\approx(Cs)^{1/2}.
\eeq

For a final presentation of the above results, we return to physical variables.
In terms of these, the survival probability falls off as
\beq
S(t)\approx\frac{2\a-1}{\gbar}\,\left(C\,\ln\frac{t}{t_0}\right)^{-1/2},
\label{stlate}
\eeq
whereas the mean mass of the surviving clusters grows as
\beq
\meansur{m}_t\approx\left(C\,t\,\ln\frac{t}{t_0}\right)^{1/2}.
\label{mtlate}
\eeq

The universality inherent in the scaling results~(\ref{slate})--(\ref{mtlate})
is unusual, because it includes the prefactor $C$,
which is itself independent of the details of the initial distribution $P(y_0)$
of square masses.
It can indeed be checked explicitly that $C$ only depends
on the {\it tail exponent}
of this distribution in the vicinity of its upper bound~$y\max$:

\begin{itemize}

\item In the bounded case ($y\max$ finite),
assuming that the distribution has a power-law behaviour
$P(y_0)\approx A(y\max-y_0)^{a-1}$ for $y_0\to y\max$,
with a tail exponent $a>0$, we obtain
\beq
C=\pi a\left(\frac{\Gamma(a+1)}{\Gamma\!\left(a+\frac32\right)}\right)^2.
\label{ca}
\eeq

\item In the unbounded case ($y\max=\infty$),
assuming that the distribution has a power-law behaviour
$P(y_0)\approx B\,y_0^{-b-1}$ for $y_0\to\infty$,
with a tail exponent $b>1/2$, we obtain
\beq
C=\pi b\left(\frac{\Gamma\!\left(b-\frac12\right)}{\Gamma(b)}\right)^2.
\label{cb}
\eeq

\end{itemize}

It is worth noticing that both results~(\ref{ca}) and~(\ref{cb})
smoothly converge to the particular value $C=\pi$ of~(\ref{cpi}),
as the tail exponents $a$ and $b$ get large:
\beq
C=\pi\left(1-\frac{3}{4a}+\cdots\right)=\pi\left(1+\frac{3}{4b}+\cdots\right).
\eeq

\section{Finite-dimensional lattices}
\label{latt}

The mean-field regime described above can be thought of
as an infinite-dimensional limit of our model.
Now, in order to include the effects of fluctuations,
we consider a finite-dimensional lattice model.
In addition to the emergence of two well-separated time scales,
the model now displays {\it metastability}:
the system gets finally trapped forever in a non-trivial attractor,
where every surviving cluster is isolated.

More specifically, clusters sit at the vertices $\n$ of a regular lattice.
Every pair of nearest neighbours interacts with a uniform coupling strength $g$.
Numerical simulations have been performed on hypercubic lattices:
the chain ($D=1$), the square lattice ($D=2$), and the cubic lattice ($D=3$).
The coordination number of these lattices is $z=2D$.
Throughout this section,
initial masses are given by the exponential distribution~(\ref{initexpo}).
Unless otherwise stated, we set $\a=1$, $g=10^{-4}$,
and $\mu$ such that~(\ref{sunexpo}) yields $S\un=0.9$.

The main focus will be the limit of weak coupling ($g\ll1$).
We therefore simplify the dynamical equations~(\ref{dsx}),
by keeping terms up to first order in $g$.
The resulting explicit equations
\beq
x'_\n=\left(\frac{2\a-1}{2}+g\sum_{\m}\left(\frac{1}{x_\m}-\a x_\m\right)
\right)x_\n-\frac{1}{x_\n},
\label{dsg}
\eeq
where $\m$ runs over the $z$ nearest neighbours of site $\n$,
are solved numerically by means of a standard first-order scheme.

\subsection{Two-step dynamics}

In the weak-coupling regime, the dynamics generated by~(\ref{dsg})
again consists of two successive well-separated stages.
As before, fast individual dynamics are exhibited in Stage~I,
while Stage~II is the arena for slow collective dynamics.
The effects of going beyond mean field are only palpable
in the latter stage, since interactions are irrelevant in Stage~I.

\subsubsection*{Stage~I: Fast individual dynamics}

There is little new here with respect to the mean-field regime.
The mass of each cluster evolves as if it were isolated, as before.
The survival probability $S(s)$ decays rather fast
from $S(0)=1$ to its plateau value $S\un$ of~(\ref{s1}).

\subsubsection*{Stage~II: Slow collective dynamics}

The slow collective dynamics throughout Stage~II
is now very different from the mean-field regime.
The survival probability $S(s)$ indeed decays
from its plateau value $S\un$ to a non-trivial limiting value $S\infy$,
because of metastability, as will be shown below.

\begin{figure}[htb]
\begin{center}
\includegraphics[angle=90,width=.6\linewidth]{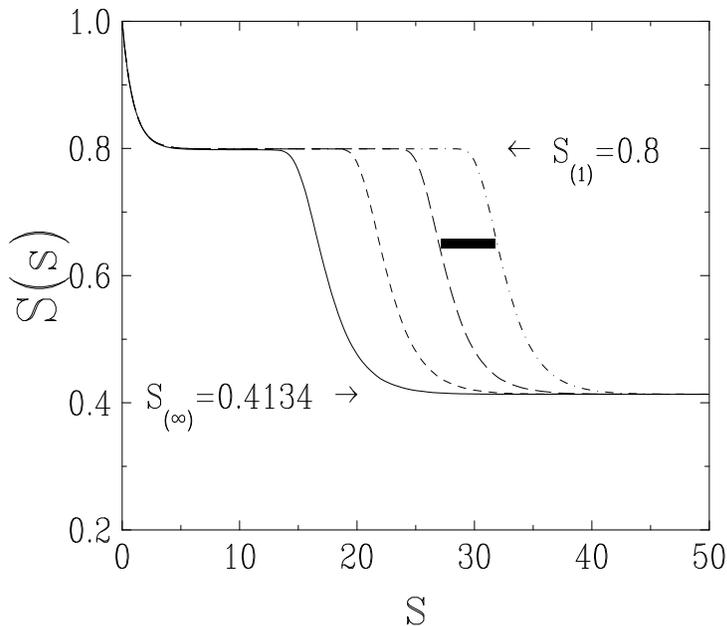}
\caption{\small
Plot of the survival probability $S(s)$ on the chain with $S\un=0.8$.
Left to right:
Full line: $g=10^{-3}$.
Dashed line: $g=10^{-4}$.
Long-dashed line: $g=10^{-5}$.
Dash-dotted line: $g=10^{-6}$.
The thick bar has length $2\,\ln 10=4.605$~(see text).}
\label{figc}
\end{center}
\end{figure}

The collective dynamics throughout Stage~II
is very slow in the weak-coupling regime.
Consider now~(\ref{dsg}) for two neighbouring clusters $\n$ and $\m$
which have both survived Stage~I.
The contribution of cluster $\m$ to the large parenthesis in the
right-hand side of~(\ref{dsg}) is proportional to~$\a gx_\m$.
In the absence of coupling,
we have $x_\m\sim\e^{(2\a-1)s/2}$, by virtue of~(\ref{yone}).
The characteristic time scale of Stage~II
is reached when the product~$gx_\m$ becomes of order unity.
It reads therefore
\beq
s_c\approx\frac{2}{2\a-1}\,\ln\frac{1}{g},
\label{sc}
\eeq
i.e.,
\beq
t_c\sim t_0\,g^{-2/(2\a-1)}.
\eeq
The separation of time scales between the fast individual
and the slow collective dynamics
is therefore again parametrically large in the weak-coupling limit,
although the divergence of the collective time scale is much less pronounced
than in the mean-field limit~[see~(\ref{scmf})].
The glassiness of the dynamics, with its manifest two-step relaxation,
is illustrated in Figure~\ref{figc}.
This figure
shows a plot of the decay of the survival probability $S(s)$ in one dimension.
The dynamical equations~(\ref{dsg})
have been integrated numerically for a chain of $10^6$ clusters,
until every surviving cluster is isolated (see below).
Both stages of the dynamics appear clearly on the plot,
as well as the expected plateau value $S\un=0.8$,
and the occurrence of a non-trivial limit survival probability
$S\infy\approx0.4134$.
Each curve corresponds to an interaction strength $g$
a decade apart from its neighbour.
It is accordingly shifted by~$2\,\ln 10$ (thick bar),
in accord with the estimate~(\ref{sc}).

At the end of Stage~II of the dynamics, i.e.,
in practice after a very long time,
the system is left in a non-trivial {\it attractor},
which consists in a pattern where each cluster is {\it isolated}:
all its first neighbours are dead, and is therefore a {\it survivor}:
it survives and keeps growing forever.
In the following, we shall call these attractors {\it metastable states},
in analogy with a variety of statistical-mechanical systems,
where metastable states have been identified
under various names in different contexts:
valleys~\cite{ks}, pure states~\cite{tap,ktw},
inherent structures~\cite{sw}, quasi-states~\cite{fv}.
The common feature of metastable states in all these situations
is that their number~$\N$
generically grows exponentially with the system size (number of sites) $N$:
\beq
\N\sim\exp(N\Sigma).
\eeq
The quantity $\Sigma$ is usually referred to as the configurational entropy,
or complexity.

The rest of this section is devoted to various characteristics
of these attractors, such as their density
(equal to the limit survival probability), spatial patterns,
spatial correlations, and mass distribution of survivors.

\subsection{Limit survival probability}

The limit survival probability $S\infy$,
already emphasised in Figure~\ref{figc},
is just the density of a typical attractor,
i.e., the fraction of the initial clusters which survive forever
and take part in the attractor.
The limit survival probability obeys the inequalities
\beq
S\infy\le S\un,\qquad S\infy\le1/2.
\eeq
The first inequality expresses that clusters can only disappear:
the difference $1-S\un$ (resp.~$S\un-S\infy$) is the fraction of clusters
which die out during Stage~I (resp.~Stage~II).
The second inequality is a consequence of the fact that
each surviving cluster is isolated.
The densest configuration of lattice sites
obeying this condition consists in occupying all the sites
of either of the two sublattices, whose density is exactly $1/2$.
This value $1/2$ of the highest density holds for the large family
of so-called {\it bipartite} lattices,
which includes hypercubic lattices (chain, square lattice, cubic lattice, ...).
It is, however, not universal,
and would e.g.~be only $1/3$ for the triangular lattice.

In a given class of initial mass distributions,
the limit survival probability $S\infy$
is a monotonically increasing function of the plateau value $S\un$,
starting from $S\infy=0$ for $S\un=0$,
and going to a non-trivial maximum value $S\infymax<1/2$
in the $S\un\to1$ limit.

In the regime where $S\un$ is small,
it can be shown that $S\infy$ is also small,
and that it depends on $S\un$ alone.
To do so, let us introduce the concept of supercluster.
In analogy with a percolation cluster,
a supercluster is defined as a set of $k\ge1$ connected clusters
which have survived Stage~I,
and such that all their neighbours have disappeared during Stage~I.
The fate of superclusters depends on their size $k$ as follows.

\begin{itemize}

\item[$\star$] $k=1$:
If a supercluster consists of a single isolated cluster,
it evolves in Stage~II according to the dynamics of Section~\ref{one}:
it is a survivor, because its reduced square mass exceeds
the threshold $y^\star$ of~(\ref{ystar}).
For independent of initial masses,
a supercluster with $k=1$ occurs with density $p_1=S\un(1-S\un)^{2D}$.

\item[$\star$] $k=2$:
If a supercluster consists of a pair of neighbouring clusters
(represented as~\pl\pl)
both clusters evolve according to the dynamics of Section~\ref{twoun}:
the smaller dies out, while the larger is a survivor.
We are thus left with~{\pl\vi} or~{\vi\pl} in the late stages of the dynamics.
Such an event takes place with density $p_2=S\un^2(1-S\un)^{2(2D-1)}$.

\item[$\star$] $k\ge3$:
If three or more surviving clusters form a supercluster,
they may a priori have more than one possible fate.
Consider for instance a linear supercluster of three clusters (\pl\pl\pl).
If the middle one disappears first (\pl\vi\pl),
the two end ones are isolated, and both will be survivors.
If one of the end ones disappears first (e.g.~\pl\pl\vi),
the other two form an interacting pair,
and only the larger of those two will survive forever (e.g.~\pl\vi\vi).
The pattern of the survivors,
and even their number, therefore cannot be predicted a priori.

\end{itemize}

The above enumeration implies $S\infy=p_1+p_2/2+\cdots$,
where the dots stand for the unknown contribution of superclusters
with $k\ge3$.
As $p_1\sim S\un$, $p_2\sim S\un^2$, and so on, we are left with the expansion
\beq
S\infy=S\un-D\,S\un^2+\cdots
\label{expan}
\eeq
The dependence of $S\infy$ on details of the initial mass distribution
at fixed $S\un$ therefore only appears at order $S\un^3$.

\begin{figure}[htb]
\begin{center}
\includegraphics[angle=90,width=.6\linewidth]{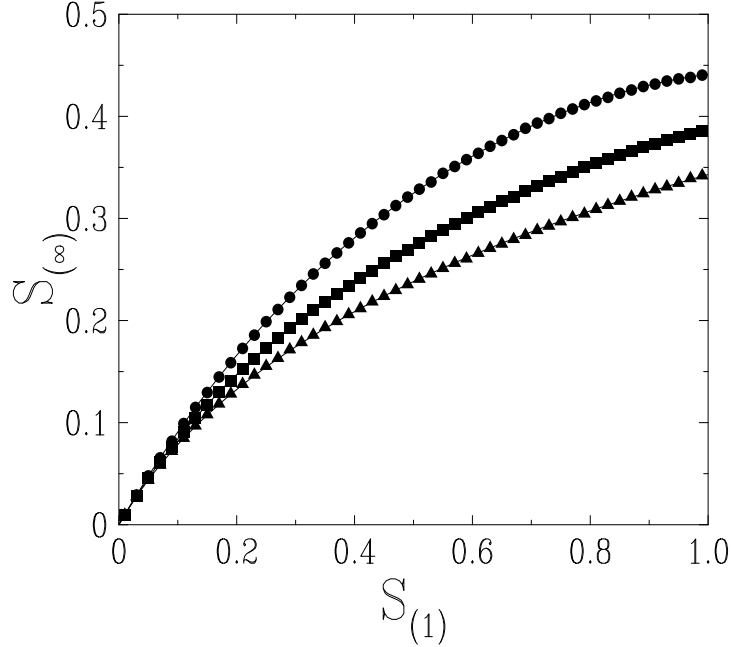}
\caption{\small
Plot of the limit survival probability $S\infy$
against the plateau survival probability~$S\un$.
Top to bottom: one dimension (circles), two dimensions (squares),
three dimensions (triangles).
Full lines (hardly visible through symbols): rational fits
based on [2/2] Pad\'e approximants
incorporating both terms of the expansion~(\ref{expan}).}
\label{figd}
\end{center}
\end{figure}

In the converse limit $S\un\to1$, the limit survival probability
reaches a non-trivial maximum value $S\infymax<1/2$,
which depends very weakly on the mass distribution.
For instance, in one dimension one has
$S\infymax\approx0.441$ for an exponential distribution
and $S\infymax\approx0.446$ for a uniform distribution.
Figure~\ref{figd} shows a plot of the limit survival probability $S\infy$
(fraction of clusters that survive both fast and slow dynamics)
against the plateau survival probability $S\un$
(fraction of clusters that survive the fast dynamics of Stage~I),
for an exponential mass distribution in one, two, and three dimensions.

\subsection{Spatial patterns of attractors}

We have seen already that the system generically ends up
trapped in an attractor, or metastable state, where each cluster is isolated,
i.e., surrounded by empty sites, and therefore grows forever.
The spatial patterns generated by survivors
are therefore absolutely fixed once they are created by the dynamics.

We now turn to a descriptive investigation of these patterns.
One of our main rationales for this exploration of spatial patterns
derives from the cosmological model which was at the origin of the present
model~\cite{I}.
In that context, it was of interest to obtain both the mass distribution
of black holes and their spatial pattern.

For ease of visualisation, we consider a square lattice.
In the limit of highest density ($S\infy=1/2$),
there are only two possible `ground-state' configurations of the system:
one where the first sublattice is full of survivors, while the second is empty,
and vice-versa.
In this limit,
patterns of surviving clusters are therefore perfect checkerboards.

In order to describe the patterns of attractors,
we are led to introduce at every site~$\n$,
with integer co-ordinates $(n_1,n_2,\dots,n_D)$,
both the {\it survival index}
\beq
\s_\n=\left\{\matrix{
1\quad\hbox{if the cluster at site~$\n$ is a survivor,}\hfill\cr
0\quad\hbox{else,}\hfill}\right.
\eeq
and the {\it checkerboard index}
\beq
\phi_\n=(-1)^{\s_\n+n_1+\cdots+n_D}.
\eeq
The survival index depicts very simply the pattern of surviving clusters
surrounded by empty sites.
The checkerboard index, on the other hand, represents,
for each site, the local choice of one of the two symmetry-related
`ground states', i.e., of one of the two sublattices.
This is easiest to understand using a one-dimensional example:
the two ground states are $+-+-+\cdots$ or $-+-+-\cdots$
All the $\phi_n$ are equal to $-1$ in the first pattern,
and equal to $+1$ in the second pattern.
The checkerboard index $\phi_\n$ thus classifies
each site according to the particular ground state selected locally
at this site.

If the initial masses are large enough, so that the plateau
survival probability $S\un$ after Stage~I is close to unity,
the limit survival probability $S\infy$ is not far
from its `ideal' highest value of~$1/2$.
In this regime, attractors clearly exhibit a local checkerboard structure,
as well as frozen-in defects with respect to a perfect checkerboard.
The random structure of defects is entirely inherited
from the random distribution of initial masses,
because the dynamics is deterministic.

\begin{figure}[htb]
\begin{center}
\includegraphics[angle=0,width=.35\linewidth]{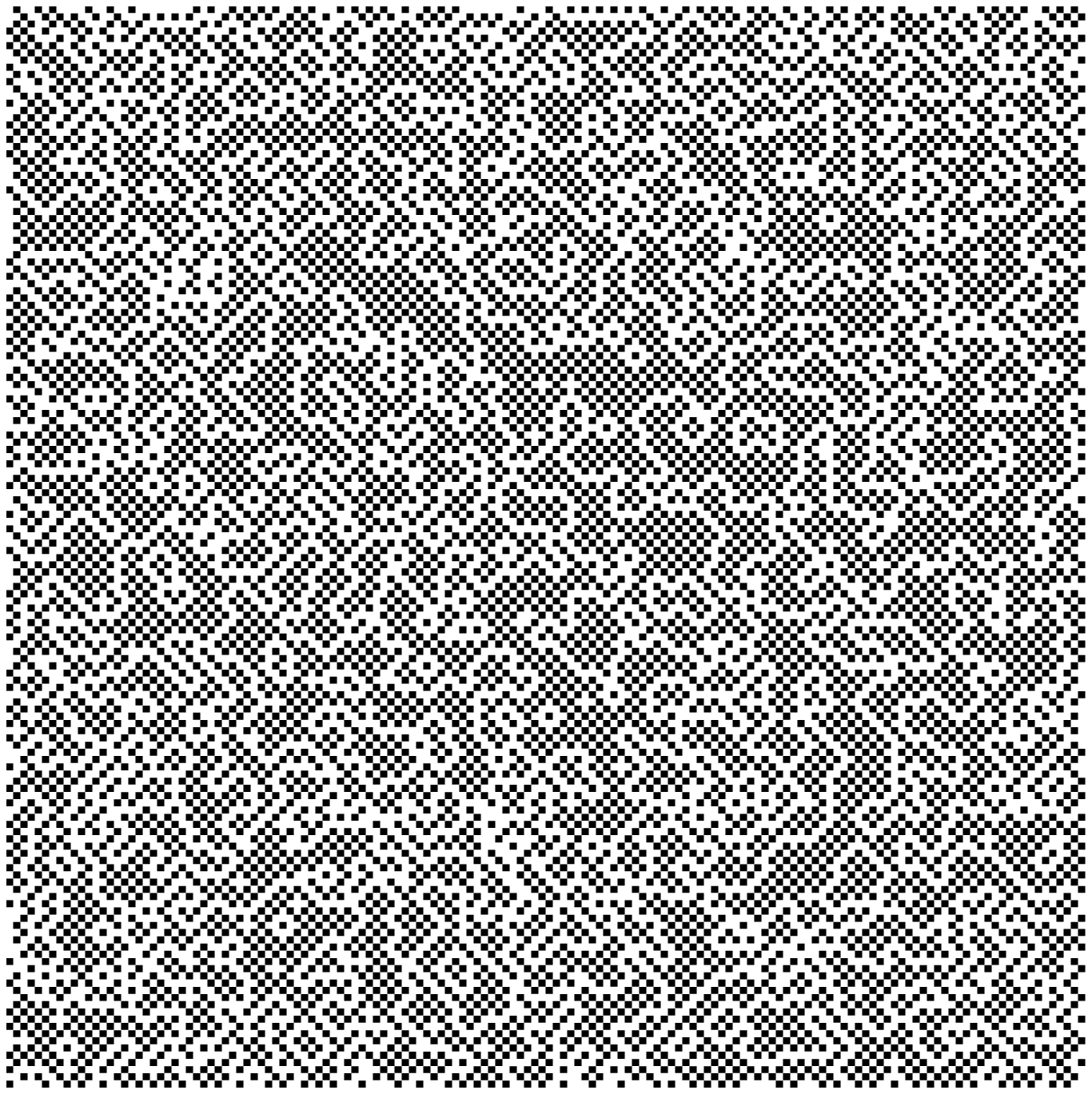}
{\hskip 1mm}
\includegraphics[angle=0,width=.35\linewidth]{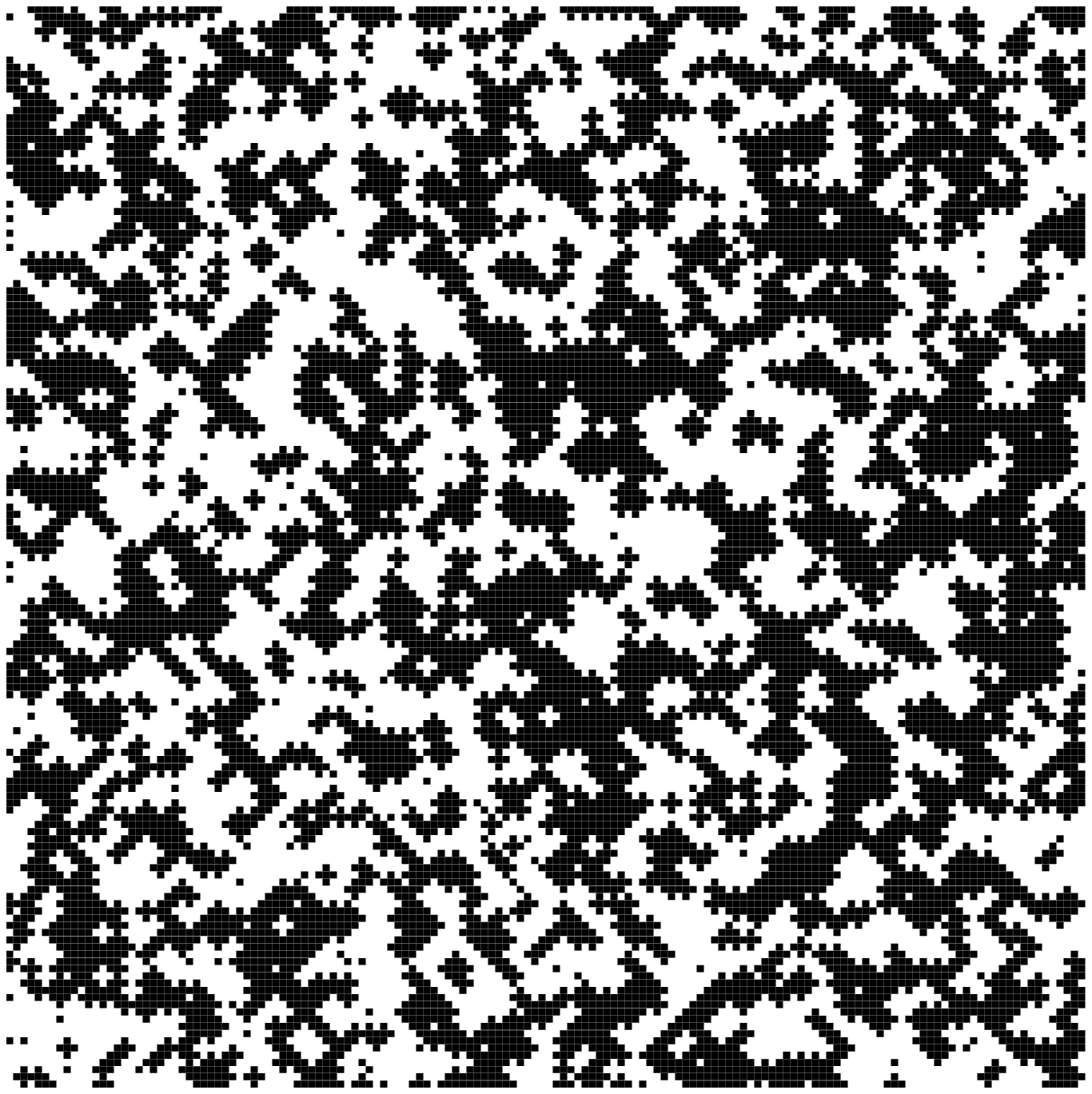}
\vskip 3mm

\includegraphics[angle=0,width=.35\linewidth]{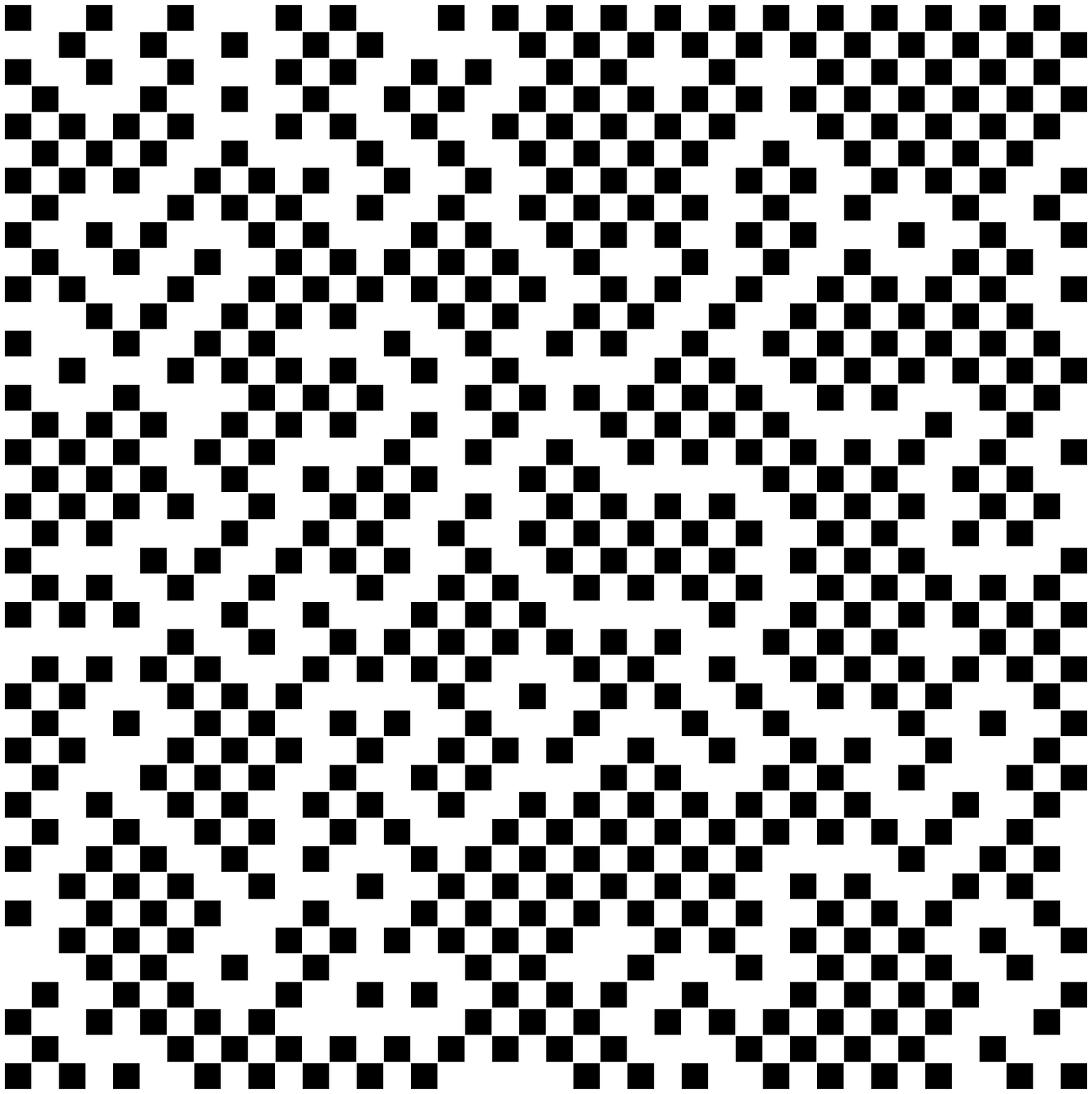}
{\hskip 1mm}
\includegraphics[angle=0,width=.35\linewidth]{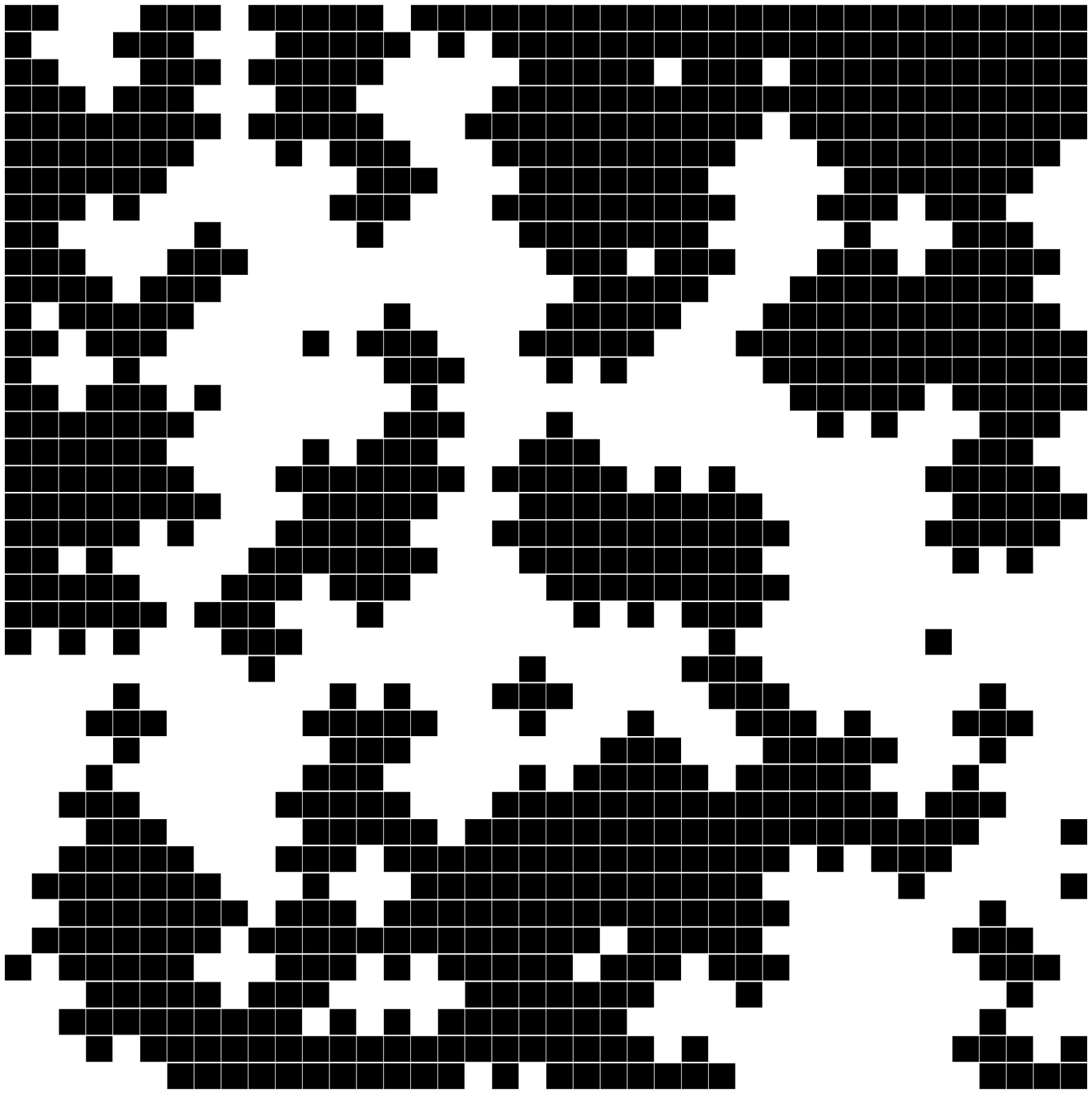}
\caption{\small Two complementary representations
of a typical pattern of surviving clusters on the square lattice,
with $S\un=0.9$, so that $S\infy\approx0.371$.
Left: Map of the survival index.
Black (resp.~white) squares represent $\s_\n=1$ (resp.~$\s_\n=0$),
i.e., surviving (resp.~dead) sites.
Right: Map of the checkerboard index.
Black (resp.~white) squares represent $\phi_\n=+1$ (resp.~$\phi_\n=-1$).
Top panels show a $150^2$ sample.
Bottom panels show an enlargement of a $40^2$ region
near the centre of the sample.}
\label{fige}
\end{center}
\end{figure}

Figure~\ref{fige} shows a map of the survival index
and of the checkerboard index
for the same attractor of a $150^2$ sample of the square lattice.
This attractor has a density $S\infy\approx0.371$.
For greater clarity, we also zoom into a part of size $40^2$,
in order to show better the correspondence between
patterns of the survival and checkerboard indices.
The frozen-in defects cause what appear to be little rivulets of voids
which surround patches of perfect checkerboard.
These islands of checkerboard are represented by black
or white patches in the lower right-hand figure, depending on their parity,
which is clearly visible from a comparison of the two figures.

\subsection{Spatial correlations}

The main consequence of short-range interactions
is the generation of correlations between clusters.
In our model, a study of such correlations is especially meaningful
in the long-time limit, when all the clusters which are still present
are in fact survivors.
In this regime, neighbouring sites are fully anticorrelated,
because each survivor is surrounded by voids.
However, at least close to the limit $S\infy=1/2$,
the next-nearest neighbours of a surviving cluster
are expected to contain another survivor with high probability.
Also, most survivors at late times should be quite massive,
as they have both survived Stage~I and then Stage~II.
We may thus expect that survival and mass correlations
exhibit a rather similar dependence on the distance.
These expectations are borne out by the following detailed study of correlation
functions.

Let us introduce the two-point correlation functions
of the survival index and of the reduced mass at separation $\n$:
\beq
C_\s(\n)=\frac{\mean{\s_\0\s_\n}}{S\infy},\qquad
C_x(\n)=\frac{\mean{x_\0x_\n}}{\mean{x^2}},
\eeq
where $\mean{\dots}$ stands for a normalised ensemble average
at a very late stage of the dynamics.
These correlations are normalised so as to have
$C_\s(\0)=C_x(\0)=1$, whereas $C_\s(\evec)=C_x(\evec)=0$,
where $\evec$ is any unit vector of the lattice.
Their disconnected parts read
\beq
C_\s(\infty)=\lim_{\abs{\n}\to\infty}C_\s(\n)=S\infy,\qquad
C_x(\infty)=\lim_{\abs{\n}\to\infty}C_x(\n)=\frac{\mean{x}^2}{\mean{x^2}}.
\label{disco}
\eeq

It is worth noticing that the correlations of the checkerboard index
are not independent of those of the survival index.
We have indeed:
\beq
\mean{\phi_\0\phi_\n}=(-1)^{n_1+\cdots+n_D}(4S\infy C_\s(\n)-4S\infy+1).
\eeq
The existence of such an identity is quite natural,
although it may seem surprising at first sight.
Indeed the checkerboard index is only a different bookkeeping method
for the same data on the positions of the survivors.

\begin{figure}[htb]
\begin{center}
\includegraphics[angle=90,width=.6\linewidth]{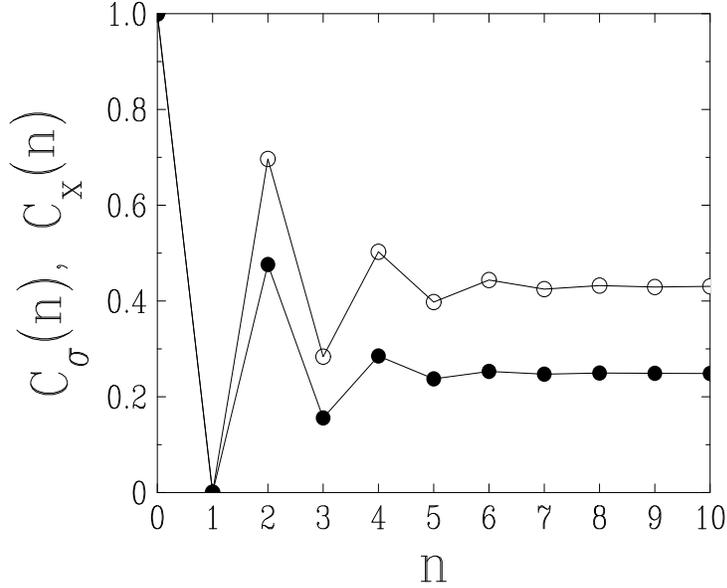}
\caption{\small Plot of the correlation functions
against the distance $n$ along the chain.
Empty symbols: correlation $C_\s(n)$ of the survival index.
Full symbols: correlation $C_x(n)$ of the reduced mass.}
\label{figf}
\end{center}
\end{figure}

\begin{figure}[htb]
\begin{center}
\includegraphics[angle=90,width=.48\linewidth]{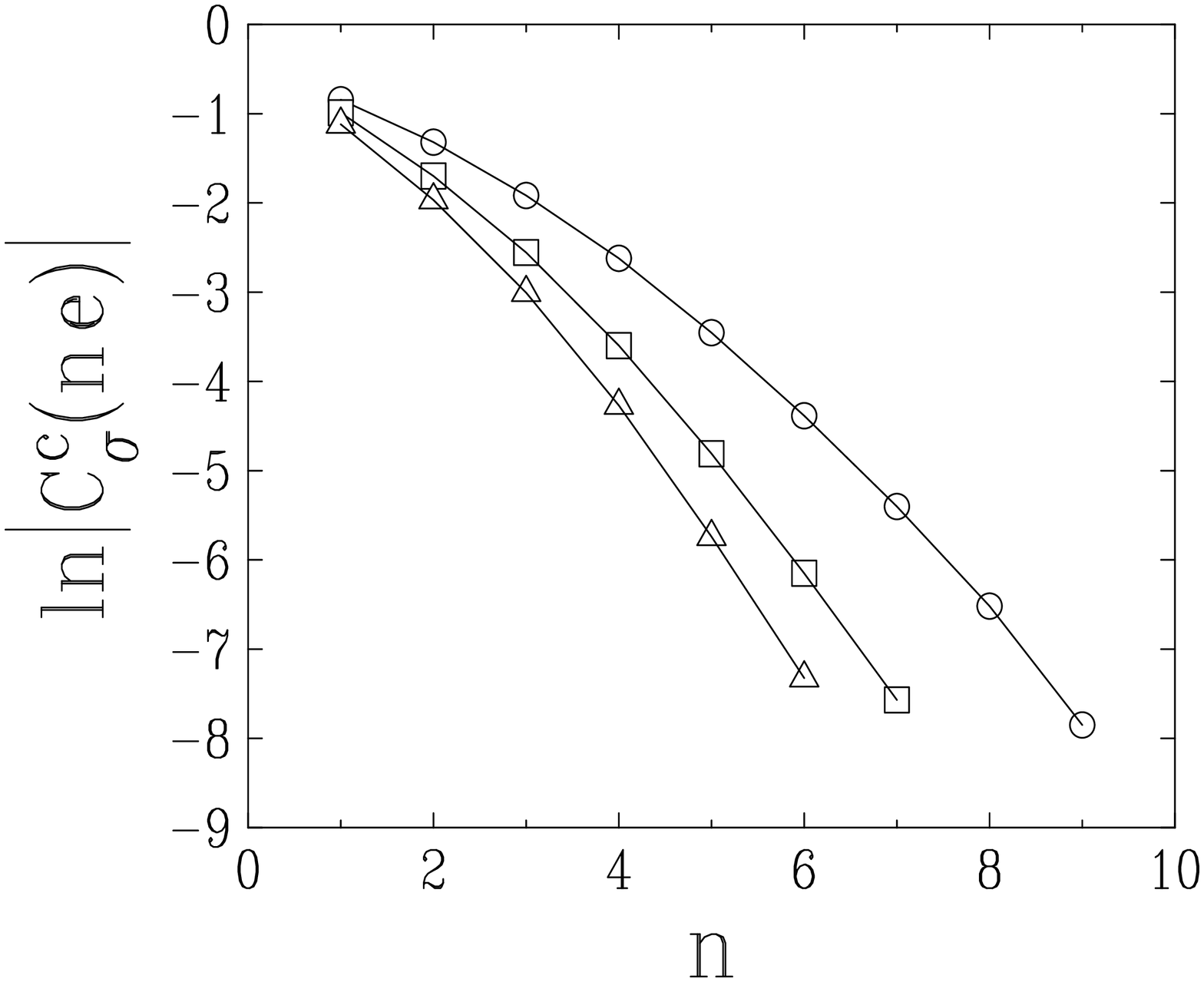}
\includegraphics[angle=90,width=.48\linewidth]{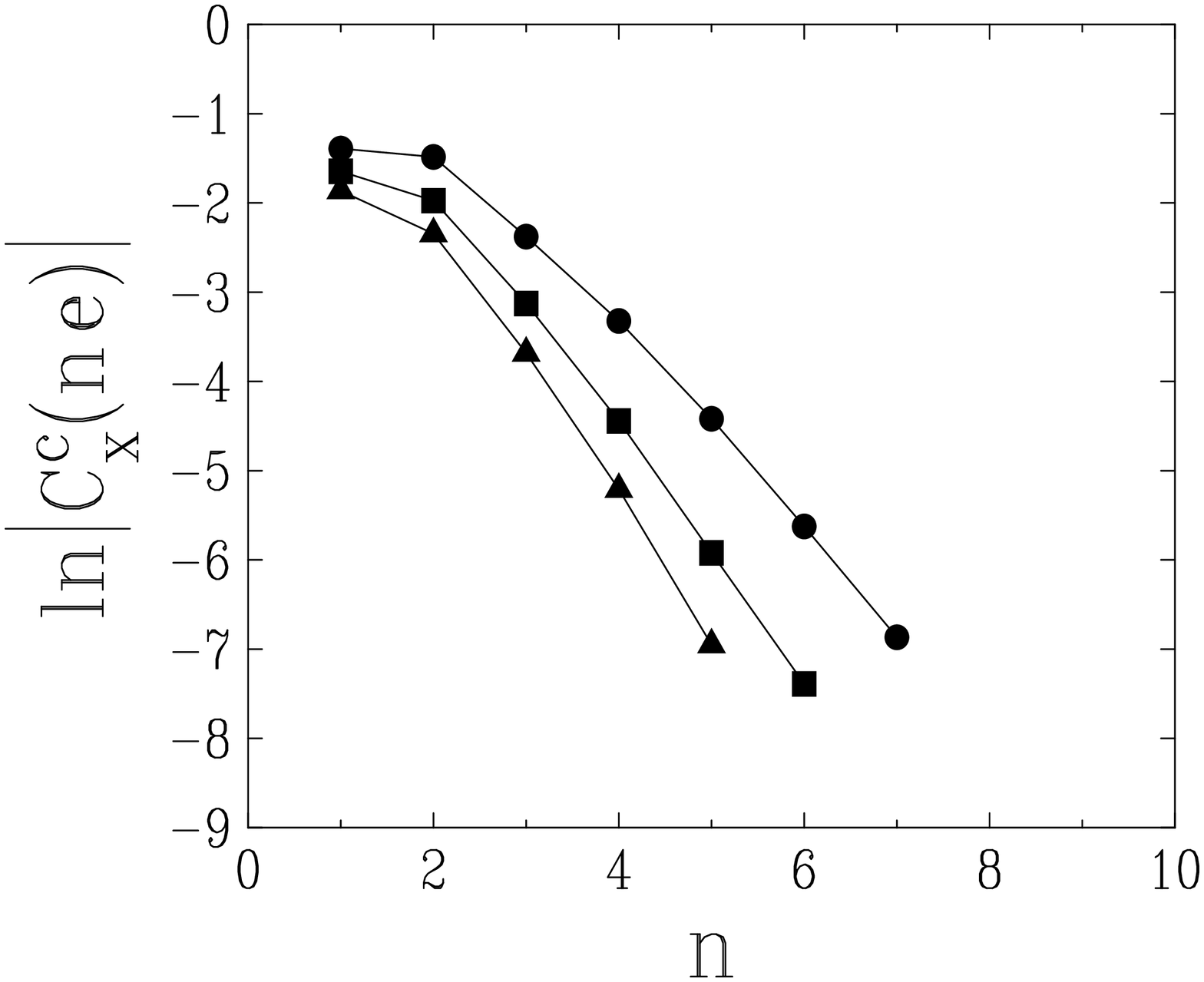}
\caption{\small Logarithmic plot of the absolute
connected on-axis correlation functions, against distance $n$.
Left: survival correlation.
Right: mass correlation.
Top to bottom: one dimension (circles), two dimensions (squares),
three dimensions (triangles).}
\label{figgh}
\end{center}
\end{figure}

Figure~\ref{figf} shows a plot of the correlation functions
$C_\s(n)$ and $C_x(n)$ against distance, in one dimension.
Both correlation functions exhibit a fast oscillatory convergence
toward their disconnected parts.
Similar features are observed for the on-axis correlations
$C_\s(n\evec)$ and $C_x(n\evec)$ on the square and cubic lattices.
Figure~\ref{figgh} shows
a plot of the logarithm of the absolute on-axis connected correlations
$C_\s^c(n\evec)=C_\s(n\evec)-C_\s(\infty)$
and $C_x^c(n\evec)=C_x(n\evec)-C_x(\infty)$, against distance,
in one, two, and three dimensions.
The connected correlations are observed to fall off very fast to zero,
so fast that it is hard to fit the precise form of their asymptotic decay.
Neither a conventional exponential fall-off
nor a more exotic super-exponential behaviour
can be ruled out from the available data.
A more accurate investigation of this point
will form the subject of future investigations.
It is worth recalling that,
in the context of zero-temperature dynamics of Ising spin chains,
the super-exponential fall-off of correlations in metastable states
has been emphasised as a signature of the generation
of a non-trivial measure on the space of attractors~\cite{smedt},
i.e., loosely speaking,
of the violation of Edwards' flatness hypothesis~\cite{edwards}.

\subsection{Mass distribution of survivors}

\begin{figure}[htb]
\begin{center}
\includegraphics[angle=90,width=.6\linewidth]{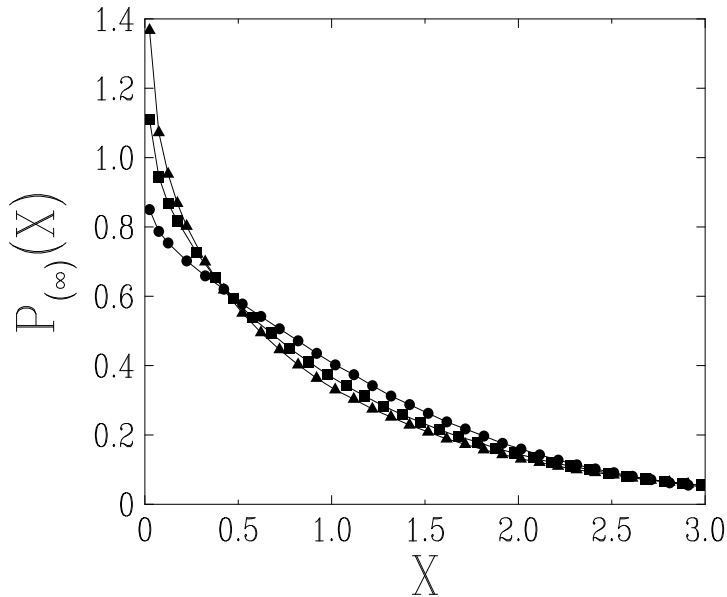}
\caption{\small
Plot of the limit distribution $P\infy(X)$
of the rescaled mass $X$ of surviving clusters.
Circles: one dimension.
Squares: two dimensions.
Triangles: three dimensions.}
\label{figi}
\end{center}
\end{figure}

To conclude this section, we examine the mass distribution of the survivors.
In the late stages of the dynamics, when every cluster is isolated,
its reduced mass grows according to Section~3, i.e., $x\sim\e^{(2\a-1)s/2}$.
The mean reduced mass $\meansur{x}$ of the surviving clusters
therefore exactly follows the same growth law.
Hence it is natural to measure cluster masses with respect to their mean,
and to introduce the ratios
\beq
X_\n=\frac{x_\n}{\meansur{x}}=\frac{m_\n}{\meansur{m}}.
\eeq
All the rescaled variables $X_\n$ are independent of time
in the very late stages of the dynamics.
They are therefore expected to have a well-defined
limit probability distribution $P\infy(X)$.
The first moment of this distribution is identically $\meansur{X}=1$,
whereas~(\ref{disco}) implies
\beq
\meansur{X^2}
=\frac{\meansur{x^2}}{\meansur{x}^2}
=S\infy\frac{\mean{x^2}}{\mean{x}^2}
=\frac{C_\s(\infty)}{C_x(\infty)}.
\eeq
We have measured the values $\meansur{X^2}\approx1.73$,
$\meansur{X^2}\approx1.93$, and $\meansur{X^2}\approx2.10$,
respectively in one, two, and three dimensions.
Figure~\ref{figi} shows a plot of the whole rescaled distribution
$P\infy(X)$ of the masses of survivors in these three cases.
This distribution is observed to be both rather structureless
and weakly dependent on dimensionality.

The cosmological origins of the present model
were our main motivation for looking at the mass distribution of survivors.
In that context, survivors would be the descendants of primordial black holes,
which would form part of the dark matter in the Universe.
The above results suggest that the population of survivors
is essentially given by a single mass scale,
growing as the mass of a single isolated cluster,
whereas the superimposed cluster-to-cluster fluctuations
are described by the harmless distribution $P\infy(X)$.
This qualitative picture clearly ignores any cosmological details,
but may still be of general interest from the viewpoint
of using statistical-mechanical methods to probe such issues.

\section{Discussion}
\label{discussion}

In the above we have presented and investigated in detail
the many facets of a novel and very rich model
at the interface between non-linear dynamics
and non-equilibrium statistical mechanics.
Despite its origins in a rather exotic context,
namely accretion dynamics of black holes in a brane world~\cite{archan,I},
the present model is of potential interest in many other situations,
at least as far as qualitative features are concerned.

As recalled in the beginning of the Introduction,
the premise of equilibration generally holds in most physical instances.
The rare exceptions to this principle
include classical systems with long-range forces,
with the noticeable example of the large-scale structure of the Universe.
Similar instabilities, where tiny initial differences get amplified forever,
are also met in other sciences,
with one well-known example being the {\it rich-get-richer} principle
in economics~\cite{simon}.
It has been realised more recently that non-equilibrium
statistical-mechanical models may exhibit a similar phenomenon,
where a single microscopic state acquires a large population
by virtue of a condensation transition, even in one dimension.
The appearance of such a condensate can be viewed as a classical
and non-equilibrium analogue of Bose-Einstein condensation~\cite{cond}.
The scenario of {\it survival of the biggest} arising from
our model can be viewed as an extreme example of this instability,
where the condensate ends up containing the entire mass.
We reiterate that the model is non-conserving, in the sense that the final
cluster becomes eventually more massive than all the earlier ones put together.
The physical reason for this is that the interaction term
derives from a radiation field in its cosmological incarnation~\cite{I},
which can be regarded as a mass reservoir.

The present model also has many other specific features
of interest in each of the geometries considered; we address each one in turn.

First, for a finite assembly of coupled clusters,
the model provides an interesting example of a deterministic dynamical system
describing the evolution of competing agents.
This class of problems has been studied at length in biophysics,
one particularly well-known example being the Lotka-Volterra system
in population dynamics as described by predator-prey models~\cite{murray}.
The present model exhibits a whole variety of types of trajectories
in the transient regime, encoded in Phases~I to~IV
of the phase diagram shown in Figure~\ref{figa}.
Another distinguishing feature of our model is that only one survivor
remains after a sufficiently long time, leading to the description
of this model as a {\it winner-takes-all} type of model,
despite the lack of a conservation law.

Next, the thermodynamical mean-field limit of our model
clearly shows many features of glassy behaviour.
The evolution consists of two successive stages:
a fast individual dynamics in Stage~I,
followed by a slow collective dynamics in Stage~II.
In the weak-coupling regime, the characteristic reduced time scale $s_c$
of the slow dynamics diverges as~$1/\gbar^2$ [see~(\ref{scmf})];
the two time scales are well-separated
in a way which is very reminiscent of the $\a$ and $\beta$ relaxations
observed in most glassy systems~\cite{glassyrefs}.
A particularly interesting feature in this context
is the universality of Stage~II asymptotics,
such as~(\ref{stlate}) or~(\ref{mtlate}),
and the more unusual universality of the prefactor $C$,
which only depends on the tail exponent of the initial mass distribution.
A further distinguishing characteristic
is that the time separation between fast and slow dynamics
in our model is simply given in terms of the coupling constant,
and becomes parametrically large in the weak-coupling regime; this
is true both within and beyond mean-field theory
[see respectively~(\ref{scmf}) and~(\ref{sc})].
In most conventional glassy systems, the separation of time scales
is governed by the appearance of a slowly growing length scale~$L(t)$,
associated with some kind of ordering.
By contrast, the glassiness in the present situation has dynamical origins:
our model has features that are similar to driven systems,
where time-scale separation arises either from
a non-zero drift velocity~$V$~\cite{ourvoter}
or a non-zero shear rate~$\gamma$~\cite{brayetal}.
Such time-scale separations become parametrically large with the divergence
(as $V\to0$ or $\gamma\to0$ for those models, and as~$g\to0$ in the present
case) of the slow time scale.

Beyond mean-field,
e.g.~on finite-dimensional lattices with nearest-neighbour couplings,
the principle of {\it survival of the biggest} only applies locally;
thus isolated clusters of dissimilar sizes are able to survive independently.
At least qualitatively, this recalls
local screening mechanisms in a variety of growth models,
including cluster aggregation in suspensions~\cite{hydrodynamic}.
In our model, however, the screening is extreme, in the sense that
growing clusters, once isolated, are survivors,
i.e., survive and keep on growing forever.
A direct consequence of this is that the model exhibits both aging
and metastability in a way that is qualitatively similar to
what is observed in the mean-field limit, even though the time scale
of the slow stage of the dynamics diverges less rapidly
in the weak-coupling regime.
Furthermore, the aging phase gets interrupted,
as the system gets eventually trapped in a metastable state
where a finite fraction $S\infy$ of the entire lattice is occupied
by isolated clusters which survive forever.
If the clusters are initially large enough,
the density $S\infy$ is only slightly below $1/2$,
so that the spatial pattern of survivors has a local checkerboard structure.
While similar checkerboard patterns have been observed
in coupled map lattices~\cite{K,k},
our attractors are distinguished by their absolute stability:
they are created irreversibly by a deterministic dynamics
from the fluctuations in the initial distribution of their initial masses.
Once created, they then survive forever.
Many questions regarding the statistics of the metastable states
thus obtained remain open.
It would be most interesting to know whether they
are generated with a uniform measure~\`a la Edwards~\cite{edwards}
in an appropriately defined ensemble, or with a highly non-trivial one,
as suggested by recent investigations
of one-dimensional spin models~\cite{smedt,franz}.

\subsubsection*{Acknowledgements}

It is a pleasure to thank Archan Majumdar for fruitful discussions,
and Hugues Chat\'e for valuable comments,
and especially for making us aware of References~\cite{K,k}.

\end{document}